\documentclass[prc,twocolumn,nofootinbib]{revtex4}
\usepackage{amsmath,amssymb,epsfig,bm,mathrsfs,feynmp,slashed,color,isotope}
\usepackage[colorlinks=true,linkcolor=blue,citecolor=blue,urlcolor=blue]{hyperref}
\usepackage{nicefrac}
\usepackage{exscale} 
\newcommand{\bea}{\begin{eqnarray}}
\def\be{\begin{eqnarray}}
\def\ee{\end{eqnarray}}
\newcommand{\eea}{\end{eqnarray}}
\newcommand{\vecn}{{\bm n}}
\newcommand{\vecp}{{\bm p}}
\newcommand{\vecq}{{\bm q}}
\newcommand{\vecv}{{\bm v}}
\newcommand{\vecE}{{\bm E}}
\newcommand{\vece}{{\bm e}}
\newcommand{\vecj}{{\bm j}}
\newcommand{\vecH}{{\bm H}}
\newcommand{\vech}{{\bm h}}

\newcommand{\vecXi}{{\bm \Xi}}
\newcommand{\ep}{\varepsilon}
\newcommand{\Tr}{{\rm Tr}}

\newcommand{\ie}{{\it i.e.}}

\def\XXint#1#2#3{{\setbox0=\hbox{$#1{#2#3}{\int}$}
     \vcenter{\hbox{$#2#3$}}\kern-.5\wd0}}

\def\apj{ApJ}%
\def\apjs{ApJS}%
\def\aap{A\&A}%
\def\mnras{MNRAS}%
\def\pra{Phys. Rev. A}%
\def\prc{Phys. Rev. C}%
\def\prd{Phys. Rev. D}%
\def\prl{Phys. Rev. Lett.}%
\def\sovast{Soviet Ast.}%
\definecolor{red}{rgb}{0.8,0,0}
\definecolor{violet}{rgb}{0.4,0,0.4}
\definecolor{green}{rgb}{0,0.5,0.0}
\definecolor{navy}{rgb}{0.0,0.0,0.6}
\definecolor{orange}{rgb}{0.8,0.2,0.0}
   
\begin{document}
\title{
Electrical conductivity of a warm neutron star crust in magnetic fields
}
\author{Arus Harutyunyan }\thanks{\tt arus@th.physik.uni-frankfurt.de}
\affiliation{Institute for Theoretical 
   Physics, J.~W.~Goethe University, D-60438 Frankfurt-Main, Germany 
}

\author{Armen Sedrakian} \thanks{\tt sedrakian@th.physik.uni-frankfurt.de }
\affiliation{Institute for Theoretical 
   Physics, J.~W.~Goethe University, D-60438 Frankfurt-Main, Germany 
}

\begin{abstract}
  We study the electrical conductivity of finite-temperature crust of
  a warm compact star which may be formed in the aftermath of a
  supernova explosion or a binary neutron star merger as well as when
  a cold neutron star is heated by accretion of material from a
  companion. We focus on the temperature-density regime where plasma
  is in the liquid state and, therefore, the conductivity is dominated
  by the electron scattering off correlated nuclei. The dynamical
  screening of this interaction is implemented in terms of the
  polarization tensor computed in the hard-thermal-loop effective
  field theory of QED plasma.  The correlations of the background
  ionic component are accounted for via a structure factor derived
  from Monte Carlo simulations of one-component plasma.  With this
  input we solve the Boltzmann kinetic equation in relaxation time
  approximation taking into account the anisotropy of transport due to
  the magnetic field. The electrical conductivity tensor is studied
  numerically as a function of temperature and density for carbon and
  iron nuclei as well as density-dependent composition of
  zero-temperature dense matter in weak equilibrium with electrons.
  We also provide accurate fit formulas to our numerical results as
  well as supplemental tables which can be used in dissipative
  magneto-hydrodynamics simulations of warm compact stars.
\end{abstract}

\maketitle

\section{Introduction}
\label{sec:introduction}

Electrical conductivity of crustal matter in neutron stars and
interiors of white dwarfs plays a central role in the astrophysical
description of these compact stars. The spectrum of problems where the
conductivity of material is important includes magnetic field decay
and internal heating, propagation of plasma waves, various
instabilities, etc. Transport in highly compressed matter has been
studied extensively in the cold regime, \ie, for temperatures $T\le 1$
MeV ($1.16\times 10^{10}$ K), which is relevant for neutron stars
several minutes to hours past their formation in a supernova event, as
well as for the interiors of white dwarfs. Initial studies of
transport in dense matter appropriate for white dwarf stars go back to
the work by Mestel and Hoyle \cite{1950PCPS...46..331M} and
Lee~\cite{1950ApJ...111..625L} in the 1950s, who computed the thermal
conductivity of the electron-ion plasma in nonrelativistic electron
regime, relevant for the radiative and thermal transport in white
dwarfs. The electrical conductivity of ultracompressed matter, where
electrons become relativistic (at zero temperature this corresponds to
density $10^6$ g cm$^{-3}$) was computed by Abrikosov in
1963~\cite{1964Abrikosov} including the regime where matter is
solid. These initial estimates were followed by a series of works in
the 1960s and
1970s~\cite{1966ApJ...146..858H,1968PhRv..170..306L,1967MNRAS.136...27M,1968ApJ...154..557I,1968PhRv..174..276L,1969ApJS...18..297H,1970ApJ...161..553S,1970ApJ...159..641C,1973A&A....28..315K},
among which the variational study of Flowers and
Itoh~\cite{1976ApJ...206..218F} provides the most comprehensive
account of transport in the solid and liquid regimes of crustal
matter, as well as of the neutron drip regime, where free neutrons
contribute to the thermal conductivity and shear viscosity of
matter. An alternative formulation in terms of Coulomb logarithm and a
detailed comparison of results of various authors was given in
Ref.~\cite{1980SvA....24..303Y}.  The regime where ions form a liquid
was studied in Ref.~\cite{1984MNRAS.209..511N}, where it was shown
that the screening of electron-ion interactions can lead to
substantial corrections in this case.  These studies were further
improved and extended in Refs.~\cite{1979ApJ...227..995E,
  1983ApJ...273..774I,1984ApJ...285..758I,
  1984ApJ...277..375M,1987Ap.....26..295S, 1993ApJ...404..268I,
  1993ApJ...418..405I,1998PhRvL..81.5556B,1999A&A...346..345P,
  2008ApJ...677..495I}, which cover a broad range of densities and
compositions appropriate for matter in white dwarfs and crusts of
neutron stars in the case of strongly degenerate electrons and
spherical nuclei.  The special case of nonspherical nuclei (pasta
phase) at the base of a neutron star crust, which may have very low
electrical resistivity, is discussed in
Ref.~\cite{2015PhRvL.114c1102H}.  The implementation of the transport
coefficients of dense matter in the dissipative magneto-hydrodynamics
(MHD) equations was discussed and the associated transport
coefficients in strong magnetic fields were computed for
the crust of a cold neutron star in the presence of magnetic fields by a
number of authors \cite{1979ApJ...227..995E,1987Ap.....26..295S}. We
confine our attention to nonquantizing fields in this work, \ie,
fields below the critical field $B\simeq 10^{14}$~G above which the
Landau quantization of electron trajectories becomes important~\cite{1999A&A...351..787P}.

The early computations of conductivity of cold neutron star matter
described above were motivated by the studies of magnetic field decay
in neutron star interiors. Recent resistive MHD simulations of
magnetized neutron stars in general
relativity~\cite{2009MNRAS.394.1727P,2013PhRvD..88d4020D,2013PhRvD..88d3011P}, including
binary magnetized neutron stars mergers and hypermassive neutron stars
formed in the post-merger phase~\cite{2015PhRvD..92h4064D} require as
an input the conductivity of warm (heated) crustal matter. In this
regime the plasma forms a liquid state of correlated ions and ionized
electrons at nonzero temperature and in nonzero magnetic field.
Such matter is also expected in proto-neutron stars newly formed in
the aftermath of supernova explosion as well as in the crusts of
neutron stars accreting material from a companion.

In this work we start addressing the necessary input for resistive MHD
simulations of such matter, specifically its electrical
conductivity. In this regime electrons are the most mobile charge
carriers and the key mechanism of the electrical conduction is the
electron scattering off the ions. There are important statistical
corrections to the free-space scattering rate: following earlier
calculations we incorporate structure factors of one-component
plasma (we do not consider mixture here); in addition we include
dynamical screening of exchanged photons which accounts for a
frequency dependent scattering rate.  The photon self-energy is
computed within the hard-thermal-loop (HTL) effective field theory
approach to polarization tensor.

The paper is organized as follows. Section~\ref{sec:regimes} discusses
the phase diagram of electron-ion plasma in the regimes of interest
for neutron stars and white dwarfs. In Sec.~\ref{sec:conduct_tensor}
we derive the electrical conductivity tensor in magnetic field
starting from the linearized Boltzmann equation for electrons.
Section~\ref{eq:relax_time} computes the matrix elements for
electron-ion scattering including the screening of the interaction in
the HTL approximation. We also discuss the input structure factor of
ions (one-component plasma). In Sec.~\ref{sec:results} we present the
numerical results for the electrical conductivity in the density,
temperature and $B$-field regimes of interest.  Our results are
summarized in Sec.~\ref{sec:conclusions}. Appendix \ref{app:A} gives
the details of the derivation of the relaxation time used in the main
text and some numerical results. We describe the computations of
polarization tensor in Appendix \ref{app:B}.

We use the natural (Gaussian) units with
$\hbar= c = k_B = k_e = 1$, $e=\sqrt{\alpha}$, 
$\alpha=1/137$ and the metric signature
$(1,-1,-1,-1)$.

\section{Physical conditions}
\label{sec:regimes}
\begin{figure}[t] 
\begin{center}
\includegraphics[width=8.0cm,keepaspectratio]{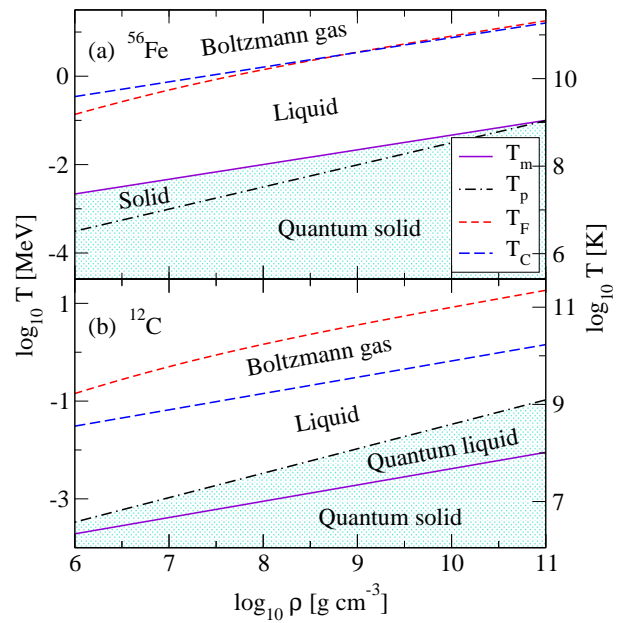}
\caption{ The temperature-density phase diagram of dense plasma
  composed of iron $\isotope[56]{Fe}$  (a) and carbon
  $\isotope[12]{C}$ (b). The electron gas degeneracy sets in
  below the Fermi temperature $T_F$ (short dashed lines). The ionic
  component solidifies below the melting temperature $T_m$ (solid
  lines), while quantum effects become important below the plasma
  temperature (dash-dotted lines).  For temperatures above $T_{\rm C}$
  (long dashed lines) the ionic component forms a Boltzmann gas. Note
  that for $\isotope[12]{C}$ the quantum effects become important
  in the portion of the phase diagram lying between the lines
  $T_p(\rho)$ and $T_m(\rho)$. The present study does not cover the
  shaded portion of the phase diagram. }
\label{fig:PhaseDiagram} 
\end{center}
\end{figure}

Matter in the interiors of white dwarfs and in the neutron star crusts
is in a plasma state -- the ions are fully ionized while free electrons
are the most mobile carriers of charge. Electron density is related to
the ion charge $Z$ by charge conservation $n_e=Zn_i$, where $n_i$ is
the number density of nuclei. Electrons to a good accuracy form
non-interacting gas which becomes degenerate below the Fermi
temperature $T_F = \varepsilon_F -m$, where the Fermi energy
$\varepsilon_F= (p_F^2+m^2)^{1/2}$, the electron Fermi momentum is
given by $p_F = (3\pi^2n_e)^{1/3}$ and $m$ is the electron mass.  The
state of ions with mass number $A$ and charge $Z$ is controlled by the
value of the Coulomb plasma parameter
\bea\label{eq:Gamma}
\Gamma=\frac{e^2 Z^2}{Ta_i}\simeq 22.73
\frac{Z^2}{T_6}\bigg(\frac{\rho_6}{A}\bigg)^{1/3},
\eea
where $e$ is the elementary charge, $T$ is the temperature,
$a_i=(4\pi n_i/3)^{-1/3}$ is the radius of the spherical volume per
ion, $T_6$ is the temperature in units $10^6$ K, and $\rho_6$ is the
density in units of $10^6$ g cm$^{-3}$.  If $\Gamma\ll 1$ or,
equivalently $T\gg T_{\rm C}\equiv Z^2e^2/a_i$, ions form
weakly coupled Boltzmann gas.  In the regime $\Gamma\ge 1$ ions are
strongly coupled and form a liquid for values of
$\Gamma\leq\Gamma_m\simeq 160$ and a lattice for
$\Gamma>\Gamma_m$. The melting temperature of the lattice associated
with $\Gamma_m$ is defined as $T_m=(Ze)^2/\Gamma_ma_i$. For
temperatures below the ion plasma temperature
\bea 
T_p = \biggl(\frac{4\pi  Z^2e^2n_i}{M }\biggl)^{1/2} ,
\eea
where $M $ is the ion mass,  the quantization of oscillations of the
lattice becomes important.  Figure \ref{fig:PhaseDiagram} shows the
temperature-density phase diagram of the crustal material in the cases
where it is composed of iron $\isotope[56]{Fe}$
 (upper panel) and carbon $\isotope[12]{C}$ (lower panel).
\begin{figure}[!] 
\begin{center}
\includegraphics[width=8.0cm,keepaspectratio]{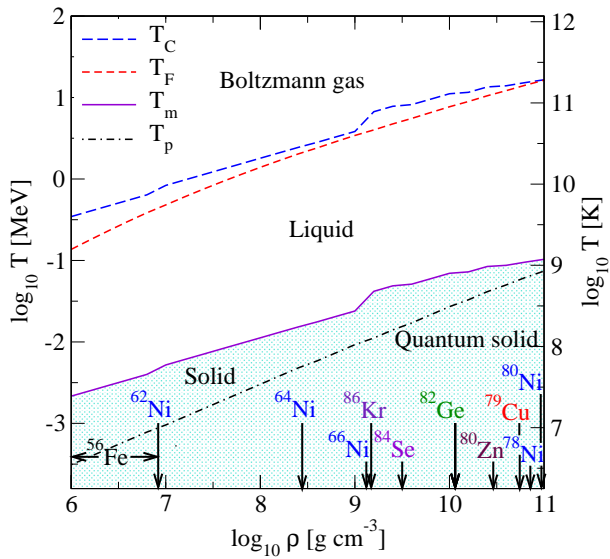}
\caption{ The temperature-density phase diagram of dense stellar
  matter in the crust of a neutron star. Various phases are labeled in
  the figure. The vertical arrows show the density at which the
  indicated element first appears. The crust composition is taken from
  Ref.~\cite{2011PhRvC..83f5810P}.  The shaded portion of the phase
  diagram indicates the regime which is not covered by our study.  }
\label{fig:Phase_Diag_Comp}
\end{center}
\end{figure}
The general structure of the phase diagram for $\isotope[56]{Fe}$
shares many common features with the phase diagram of
$\isotope[12]{C}$ however there is one important difference: as the
temperature is lowered the quantum effects become important for carbon
prior to solidification, whereas iron solidifies close to the
temperature where ionic quantum effects become important. Except for
hydrogen and perhaps helium both of which may not solidify because of
quantum zero point motions all heavier elements $Z> 2$ solidify at low
enough temperature. Figure \ref{fig:Phase_Diag_Comp} shows the same
phase diagram in the case of density-dependent crust composition
adopted from Ref.~\cite{2011PhRvC..83f5810P} where nuclei are in weak
equilibrium with electrons at zero temperature.

\section{Electrical conductivity tensor}
\label{sec:conduct_tensor}

The kinetics of electrons is described by the  Boltzmann equation for
electron distribution function 
\bea\label{eq:boltzmann1}
\frac{\partial f}{\partial t}+
\bm v\frac{\partial f}{\partial\bm r}-
e(\vecE+[\vecv\times \vecH])\frac{\partial f}
{\partial\vecp}=I[f],
\eea
where $\vecE$ and $\vecH$ are the electric and magnetic fields,
$\vecv$ is the electron velocity, $e$ is the unit charge, and $I[f]$ is
the collision integral.  In the relevant density and temperature
regime electron-ion collisions are responsible for the conductivity of
matter.\footnote{
Here we neglect the possible contribution from
  positrons, which can be sizable only in the very low density and
  high temperature matter.
}  The collision integral has the form
\bea\label{eq:collision1}
I&=&-(2\pi)^4\sum\limits_{234}|{\cal M}_{12\to 34}|^2\delta^{(4)}(p+p_2-p_3-p_4)\nonumber\\
&\times&
[f(1-f_3)g_2-f_3(1-f)g_4],
\eea
where $f=f(p)$ and $f_3=f(p_3)$ are the distribution functions of
the incoming and outgoing electron,   $g_{2,4}=g(p_{2,4})$ is the
distribution function of the ion before and after collision;  here
and below we use the short-hand notation:
$\sum\limits_{i}=\int d\vecp_i/(2\pi)^3$.
We will assume that ions form a classical ensemble in equilibrium, \ie,  their distribution
function $g(p)$ is given by the  Maxwell-Boltzmann distribution
\bea\label{eq:maxwell}
g(p)=n_i\bigg(\frac{2\pi}{MT}\bigg)^{3/2}
e^{-\beta\varepsilon},
\eea
where $\varepsilon=p^2/2M$, $M$ is the ion mass, $\beta=T^{-1}$ is the 
inverse temperature, and $n_i$ is the number density of ions.
We are interested in perturbations that introduce small deviations from
equilibrium, in which case the Boltzmann equation can be
linearized. We thus consider small perturbation around the equilibrium
Fermi-Dirac distribution function of electrons given by 
\bea\label{eq:fermi}
f^0({\varepsilon})=\frac{1}{e^{\beta 
(\varepsilon-\mu)}+1},
\eea
where $\varepsilon=\sqrt{p^2+m^2}$ and $\mu$ is the chemical potential 
and write 
\bea\label{eq:distribution}
f= f^0+\delta f,\quad \delta f=-\phi
\frac{\partial f^0}{\partial\varepsilon},
\eea
where $\delta f\ll f_0$ is a small perturbation. In the case of
electrical conduction we can keep only the last term on the
left-hand side of Eq.~\eqref{eq:boltzmann1}.  We substitute for the
electron distribution function \eqref{eq:distribution} in
Eq.~\eqref{eq:boltzmann1} and take into account the identities 
\bea\label{differmi}
\frac{\partial f^0}{\partial\vecp}=
\bm v\frac{\partial f^0}{\partial\varepsilon},
\quad\frac{\partial f^0}{\partial\varepsilon}=
-\beta f^0(1-f^0),
\eea
which follow directly from Eq.~(\ref{eq:fermi}). To linear order in
perturbation $\phi$ the Boltzmann equation reads 
\bea\label{eq:boltzmann2}
e\vecv\cdot \vecE\frac{\partial f^0}
{\partial\varepsilon}-e[\vecv\times \vecH]
\frac{\partial f^0}{\partial\varepsilon}
\frac{\partial\phi}{\partial\vecp}=-I[f],
\eea
where the collision integral in the same approximation is given by 
\bea\label{eq:collision2}
I[f]&=&-(2\pi)^4 \beta\sum\limits_{234}|{\cal M}_{12\to 34}|^2 \delta^{(4)}(p+p_2-p_3-p_4)\nonumber\\
&\times& f^0(1-f^0_3)g_2(\phi-\phi_3).
\eea
The electric field appears in the drift term of linearized Boltzmann
equation \eqref{eq:boltzmann2} at $O(1)$, whereas the
term involving magnetic field at order $O(\phi)$,  because 
$
[\vecv\times \vecH]({\partial f^0}/{\partial\vecp})\propto
[\vecv\times \vecH]\vecv=0.
$
We next specify the form of the function $\phi$ in the case of
conduction as 
\bea\label{eq:solution}
\phi=\vecp\cdot \vecXi(\varepsilon), 
\eea
which after substitution in Eqs.~\eqref{eq:boltzmann2}  and
\eqref{eq:collision2} gives
\bea\label{eq:boltzmann3}
e\vecv\cdot \left[\vecE+(\vecXi\times\vecH)\right]=-
\vecXi\cdot \vecp~\tau^{-1}(\varepsilon),
\eea
where the relaxation time is defined by 
\bea\label{eq:t_relax}
\tau^{-1}(\varepsilon)&=&(2\pi)^{-5}
\int d\omega d\bm q\int d\bm p_2|{\cal M}_{12\to 34}|^2 \frac{\bm q\cdot \bm p}{p^2}
\nonumber\\ &\times &
\delta(\varepsilon-\varepsilon_3-\omega)\delta(\varepsilon_2-\varepsilon_4+\omega) 
 g_2\frac{1-f^0_3}{1-f^0}.
\eea
In transforming the linearized collision integral we introduced a
dummy integration over energy and momentum transfers, \ie, 
$\omega = \varepsilon-\varepsilon_3$ and $\vecq = \vecp-\vecp_3$.  It
remains to express the vector $\vecXi$ describing the perturbation in
terms of physical fields. Its most general decomposition is given by
\bea\label{eq:Xi}
\vecXi=\alpha\vece+\beta\vech+\gamma[\vece\times\vech],
\eea
where $\vech \equiv \vecH/H$ and $\vece \equiv \vecE/E$ and the
coefficients $\alpha$, $\beta$, $\gamma$ are functions of the electron
energy. Substituting Eq.~\eqref{eq:Xi} in Eq.~\eqref{eq:boltzmann3}
one finds that $\alpha=-eE\tau /\varepsilon (1+\omega^2_c\tau^2)$,
$\beta/\alpha=(\omega_c\tau)^2(\vece\cdot \vech)$ and
$\gamma/\alpha=-\omega_c\tau$, where $\omega_c=eH\varepsilon^{-1}$ is
the cyclotron frequency. As a result, the most general form of the
perturbation is given by
\bea\label{eq:phi}
\phi=-\frac{e\tau}{1+(\omega_c\tau)^2}
v_i\left[\delta_{ij}-\omega_c\tau\varepsilon_{ijk}
h_k+(\omega_c\tau)^2h_ih_j\right]E_j,\nonumber\\
\eea
where the Latin indices label the components of Cartesian coordinates.
The electrical current is defined  in terms of perturbation $\phi$ as
\bea\label{eq:current}
j_i=2\int\frac{d\bm p}{(2\pi)^3}ev_i\phi \frac{\partial f^0}{\partial\varepsilon}
\eea 
and, at the same time, it is related to the conductivity tensor
$\sigma_{ij}$ by 
\bea\label{eq:ohm}
j_i=\sigma_{ij}E_j.
\eea
Substituting Eq.~\eqref{eq:phi} in Eq.~\eqref{eq:current} and
combining it with Eq.~\eqref{eq:ohm} we find for the
conductivity tensor 
\bea\label{eq:sigma1}
\sigma_{ij}=\delta_{ij}\sigma_0-\varepsilon_{ijm}h_m
\sigma_1 +h_ih_j\sigma_2,
\eea
where 
\bea\label{eq:sigma2}
\sigma_n=\frac{e^2\beta}{3\pi^2}\int_m^\infty\!\! d\varepsilon
\frac{p^3}{\varepsilon}\frac{\tau(\omega_c\tau)^n}
{1+(\omega_c\tau)^2}f^0(1-f^0),\quad n=0,1,2. \nonumber\\
\eea
The conductivity tensor has a simple form when the magnetic field is
along the $z$-direction 
\bea\label{eq:sigma3}
\hat{\sigma}=
\begin{pmatrix}
    \sigma_0 & -\sigma_1 & 0 \\
    \sigma_1 & \sigma_0 & 0 \\
    0 & 0 & \sigma
\end{pmatrix}.
\eea
Finally, note that in the absence of magnetic field 
$\vecj=\sigma\vecE$ with 
\bea\label{eq:sigma}
&&\sigma=\frac{e^2\beta}{3\pi^2}\int_m^\infty d\varepsilon
\frac{p^3}{\varepsilon}\tau f^0(1-f^0)=\sigma_0+\sigma_2.
\eea

\section{Collision integral}
\label{eq:relax_time}

We now turn to the evaluation of the collision integral, or
equivalently the relaxation time, assuming that for temperatures and
densities of interest relativistic electrons are scattered by
correlated nuclei. In free space this process is described by the
well-known Mott scattering by a Coulomb center. We use the standard
QED methods to compute the transition probability, but include in
addition the screening of the interaction by the medium in terms of
polarization tensor.

\subsection{Scattering matrix elements and relaxation time}

\begin{figure}[t]
\includegraphics[width=3cm,height=5cm,angle=90]{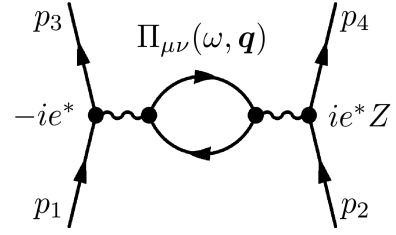}
\caption[] {Feynman diagram describing the scattering of electron of
  charge $e^*= \sqrt{4\pi}e$ off a nucleus of charge $e^*Z$
  (left and right straight arrows, respectively) via exchange of
  photon (wavy line).  The photon self-energy is given by the
  polarization tensor $\Pi_{\mu\nu}$ shown by the closed loop. Dots
  stand for QED vertices.  }
\label{fig:1}
\end{figure}

The scattering amplitude for electron scattering off a nucleus
characterized by its charge $Z$ is given by (see Fig.~\ref{fig:1}
and Appendix \ref{app:B} for details) 
\bea\label{eq:amplitude}
{\cal M}_{12\to 34}=-\frac{J_0J'_0}{q^2+\Pi'_L}+
\frac{\bm J_t\bm J'_t}{q^2-\omega^2+\Pi_T}=-{\cal M}_L+{\cal M}_T,
\nonumber\\
\eea
where
\bea\label{currents}
J^{\mu}&=&-e^*\bar{u}^{s_3}(p_3)\gamma^\mu u^s(p),\\
J'^{\mu}&=&Ze^*v'^{\mu}=Ze^*(1,\bm p'/M), 
\eea 
$e^* = \sqrt{4\pi}e$, and ${\bm J}_t, {\bm J}'_t$ are the components
of the currents transversal to $\bm q$ ($\bm p_1\equiv\bm p$,
$\bm p_2\equiv\bm p'$).  The screening of the interaction is taken
into account in terms of the longitudinal $\Pi_L(\omega,\vecq) $ and
transverse $\Pi_T(\omega, \vecq)$ components of polarization tensor,
with
$\Pi'_L(\omega,\vecq) \equiv \Pi_L(\omega,\vecq)/(1- \omega^2/q^2)$.

 The form of the matrix element
\eqref{eq:amplitude} includes thus the dynamical screening of the
electron-ion interaction due to the exchange of transverse
photons. Such separation has been employed previously in the treatment
of transport of unpaired \cite{1993PhRvD..48.2916H} and
superconducting ultrarelativistic quark
matter~\cite{2014PhRvC..90e5205A}.

Standard QED diagrammatic rules can be applied to compute the
transition probability from the diagram shown in Fig.~\ref{fig:1}. The
square of the scattering matrix element can be written as
\bea\label{probability} 
|{\cal M}_{12\to 34}|^2=|{\cal M}_L|^2+|{\cal M}_T|^2-2{\rm
  Re}{\cal M}_L {\cal M}_T^*, 
\eea 
where 
\bea\label{matrix}
|{\cal M}_L|^2&=&\frac{J_0J_0^*J_0'J_0'^*}{|q^2+\Pi'_L|^2},\\
|{\cal M}_T|^2&=&\frac{J_{i}J_{k}^*J_{ti}'J_{tk}'^*}
{|q^2-\omega^2+\Pi_T|^2},\\
{\cal M}_L {\cal M}_T^*&=&\frac{J_0J_{i}^*J_0'J_{ti}'^*}
{(q^2+\Pi'_L)(q^2-\omega^2+\Pi_T^*)}, 
\eea 
\ie, the scattering probability is split into longitudinal, transverse,
and interference contributions.  The scattering probability per unit
volume is obtained after averaging the scattering amplitude
\eqref{probability} over initial spins of electrons, summing over
final spins, and multiplying with the structure factor of ions
$S(q)$ and the square of nuclear formfactor $F^2(q)$
\bea\label{probabiliy3}
\sum\limits_{ss_3} |{\cal M}_{12\to 34}|^2 &=&
\frac{Z^2e^{*4}}{\varepsilon(\varepsilon-\omega)}~S(q) F^2(q)
\bigg\{
\frac{\varepsilon(2\varepsilon-\omega)-\bm p\cdot \bm q}
{|q^2+\Pi'_L|^2}
\nonumber\\
&+&\frac{2(\bm p_t\cdot \bm p'_t)^2+(p'_t)^2
(-\varepsilon\omega+\bm p\cdot \bm q)}
{M^2|q^2-\omega^2+\Pi_T|^2}
\nonumber\\
&-&\frac{2(2\varepsilon-\omega)(\bm p_t\cdot \bm p'_t)}
{M{\rm Re}[(q^2+\Pi'_L)(q^2-\omega^2+\Pi_T^*)]}
\bigg\}.
\eea 
Substituting the transition probability in the expression for the relaxation time 
 \eqref{eq:t_relax}, carrying our the integrations (see Appendix \ref{app:A}
for details) we finally obtain 
\bea\label{relax_final}
\tau^{-1}(\varepsilon)
&=&\frac{\pi Z^2e^4n_i}{\varepsilon p^3}
\int_{-\infty}^{\varepsilon-m} d\omega e^{-\omega/2T}
\frac{f^0(\varepsilon-\omega)}{f^0(\varepsilon)}
\nonumber\\
&\times&
\int_{q_-}^{q_+} dq(q^2-\omega^2+2\varepsilon\omega)S(q)F^2(q) \frac{1}{\sqrt{2\pi}\theta}\nonumber\\
&\times&
e^{-\omega^2/2q^2\theta^2}e^{-q^2/8MT}\bigg\{
\frac{(2\varepsilon-\omega)^2
-q^2}{|q^2+\Pi'_L|^2}\nonumber\\
&+&\theta^2
\frac{(q^2-\omega^2)[(2\varepsilon-\omega)^2
+q^2]-4m^2q^2}{q^2|q^2-\omega^2+\Pi_T|^2}\bigg\},\nonumber\\
\eea
where $\theta \equiv \sqrt{{T}/{M}}$,
$q_{\pm} =\vert \pm p+ \sqrt{p^2-(2\omega\varepsilon- \omega^2)}\vert$
and 
$\varepsilon = \sqrt{p^2+m^2}$ for non-interacting electrons. The 
contributions of longitudinal and transverse photons in 
Eq.~\eqref{relax_final} separate (first and second terms in the 
braces). 
The dynamical screening effects
contained in the transverse contribution are parametrically suppressed
by the factor $T/M$ at low temperatures and for heavy nuclei. This
contribution is clearly important in the cases where electron-electron
($e$-$e$) scattering contributes to the collision integral. This is
the case, for example, when ions form a solid lattice and, therefore,
Umklapp $e$-$e$ processes are allowed, or in the case of thermal
conduction and shear stresses when the $e$-$e$ collisions contribute
to the dissipation. 

Finally, to account for the finite size of the nuclei, we use the
simple expression for the nuclear
form factor~\cite{1984ApJ...285..758I}
\bea\label{formfactor}
F(q)=-3\frac{qr_c\cos(qr_c)-\sin(qr_c)}{(qr_c)^3}, 
\eea 
where $r_c$ is
the charge radius of the nucleus given by $r_c=1.15\, A^{1/3}$ fm (see
Appendix \ref{app:A} for numerical results).

\subsection{Recovering limiting cases}

As shown in Appendix \ref{app:A} when the ionic component of the plasma is
considered at zero temperature and nuclear recoil can be neglected the
relaxation time takes a simpler form
\bea\label{eq:relax_time3}
\tau^{-1}(\varepsilon) = \frac{\pi Z^2e^{4}n_i}{\varepsilon \, p^3}
\int_{0}^{2p} dqq^3 S(q) F^2(q)\frac{4\varepsilon^2
  -q^2}{|q^2+\Pi'_L|^2}.
\eea 
Consider now two limiting cases with respect to the temperature of the
electronic component of the plasma, the degenerate limit, \ie,
$T\ll T_F$ and the non-degenerate limit, \ie, $T\gg T_F$.  In the
zero-temperature limit Eqs.~\eqref{eq:sigma2}--\eqref{eq:sigma}
simplify via the substitution
$\partial f^0/\partial\varepsilon=- \beta f^0(1-f^0) \to
-\delta(\varepsilon-\varepsilon_F)$, \ie,
\bea\label{sigma_n_fermi}
\sigma_n
&=&\frac{e^{2}}{3\pi^2}
\frac{p_F^3}{\varepsilon_F}\tau_F\frac{(\omega_{cF}\tau_F)^n}
{1+(\omega_{cF}\tau_F)^2}
\eea
and 
\bea\label{sigmas_fermi}
\sigma =\frac{n_ee^{2}\tau_F}{\varepsilon_F},\quad 
\sigma_0=\frac{\sigma}{1+(\omega_{cF}\tau_F)^2},\quad
\sigma_1= (\omega_{cF}\tau_F)\sigma_0,\nonumber\\
\eea
where we used the expression for the electron 
Fermi momentum $p_F=(3\pi^2n_e)^{1/3}$ and defined the
relaxation time and cyclotron frequency at the Fermi energy,
$\tau^{-1}_F\equiv\tau^{-1}(\varepsilon_F)$ and 
$\omega_{cF} =eB/\varepsilon_F$ (here and below we set $B=H$ which
corresponds to permeability of matter being unity).  The first
equation in Eq.~(\ref{sigmas_fermi}) is the well-known Drude formula. From
Eq.~\eqref{eq:relax_time3} we find in the low-temperature limit
\bea\label{eq:relax_fermi}
\tau^{-1}_F=
\frac{4Ze^{4}   \varepsilon_F}{3\pi}
\int_{0}^{2p_F}
dq\, q^3\frac{ S(q) F^2(q)}{|q^2+\Pi'_L|^2}
\bigg(1-\frac{q^2}{4\varepsilon^2_F}\bigg),
\eea
where we used the charge neutrality condition $n_e=Zn_i$.
Neglecting the screening ($\Pi'_L\to 0$) and the nuclear formfactor
[$F(q)\to 1$] we obtain from Eq.\eqref{eq:relax_fermi}
\bea\label{eq::relax_Pethick}
\tau^{-1}_F = \frac{4Ze^{4}   \varepsilon_F}{3\pi}
\int_{0}^{2p_F}\frac{dq}{q}
\bigg(1-\frac{q^2}{4\varepsilon^2_F}\bigg)S(q),
\eea 
which coincides with Eqs. (9) and (11) of 
Ref.~\cite{1984MNRAS.209..511N}.

In the limit of non-degenerate electrons $f^0\ll 1$ and, therefore,
\bea\label{sigma_n_non_deg}
\sigma_n &\simeq&
\frac{e^{2}}{3\pi^2T}\int_m^\infty p^2dp
\frac{p^2}{\varepsilon^2}\frac{\tau(\omega_c\tau)^n}
{1+(\omega_c\tau)^2}f^0\nonumber\\
&=&\frac{n_ee^{2}}{3T}<v^2\frac{\tau(\omega_c\tau)^n}
{1+(\omega_c\tau)^2}>,
\eea
where the quantities in the brackets are taken at some average energy
$\bar{\varepsilon}\sim T$, which can be identified with the average
thermal energy of a particle (electron) in the Boltzmann limit. {We
  recall that the average of an energy-dependent quantity
  $F(\varepsilon)$ is defined as
\bea\label{average}
<F(\varepsilon)>=\frac{2}{n_e}\int \!\!
\frac{d\bm p}{(2\pi)^3}F(\varepsilon)f^0(\varepsilon),
\eea
where a factor 2 arises from spin of electrons.}  In
Eq. \eqref{sigma_n_non_deg} we can further replace
$v^2/3T \to 1/\bar \varepsilon$ , consequently
\bea\label{sigmas_non_deg}
\sigma \simeq \frac{n_ee^{2}\bar{\tau}}{\bar{\varepsilon}}, \quad 
\sigma_0\simeq\frac{\sigma}{1+(\bar{\omega}_c\bar{\tau})^2},\quad
\sigma_1\simeq\frac{\bar{\omega}_c\bar{\tau}\sigma}
{1+(\bar{\omega}_c\bar{\tau})^2},
\eea
where $\bar{\omega}_c=eB/ \bar{\varepsilon}$,
$\bar{\tau}=\tau(\bar{\varepsilon})$.  Thus, the formulas in both
strongly degenerate and non-degenerate regimes have the same form, but
different characteristic energy scale, which is $\varepsilon_F$ in the
degenerate regime and $\bar{\varepsilon}\simeq 3T$ in the non-degenerate,
ultrarelativistic regime. In the case where $T\sim m$  we use
$\bar{\varepsilon}= 3T/2 + \sqrt{(3T/2)^2+m^2} $, which arises
from the  condition $\bar v^2\bar \varepsilon =3T$, where
$\bar v$ is the mean velocity.

\begin{figure}[t] 
\begin{center}
\includegraphics[width=7.0cm,keepaspectratio]{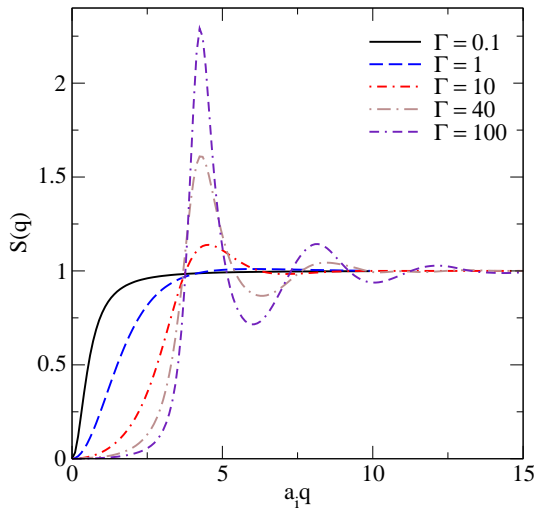}
\caption{ Dependence of the structure factor of one-component plasma
  on the magnitude of momentum transfer $q$ in units of inverse $a_i$.
  For $\Gamma\ge 2$ the structure factor is taken from Monte-Carlo
  calculations of Galam and Hansen \cite{1976PhRvA..14..816G}. For
  $\Gamma<2$ we obtain the structure factor from the analytical
  expressions provided by Tamashiro {\it et al.}~\cite{1999PhyA..268...24T}.
}
\label{fig:Sq}
\end{center}
\end{figure}

\subsection{Ion structure factor}
For the numerical computations we need to specify the ion structure
function $S(q)$. We assume that only one sort of ions exists at a
given density, so that the structure factor of one-component plasma
(OCP) can be used. These has been extensively computed using various
numerical methods. We adopt the Monte Carlo results of Galam and
Hansen~\cite{1976PhRvA..14..816G} for Coulomb OCP provided in tabular
form for $\Gamma \ge 2$
and set a two-dimension spline function in the space spanned by
the magnitude of the momentum transfer $q$ and the plasma parameter
$\Gamma$.  For small momentum transfers,  $q a_i < 1$, we use the 
formulas (A1) and (A2) of Ref.~\cite{1983ApJ...273..774I}.

In the low-$\Gamma$ regime ($\Gamma\le 2$), not covered by the
Monte Carlo results, we use the analytical (leading order) expressions
for Coulomb OCP by Tamashiro {\it et al.}~\cite{1999PhyA..268...24T} derived
using density functional methods. The dependence of the resulting
structure factors on the dimensionless parameter $a_iq$, where $a_i$
is the ion-sphere radius as defined after Eq.~\eqref{eq:Gamma}, is
shown in Fig.~\ref{fig:Sq} for various values of the plasma parameter
$\Gamma$. Note that these correlation functions were derived for
classical plasma, therefore the quantum aspects of motion of
$\isotope[12]{C}$ in the temperature regime $T_p\ge T\ge T_m$ are not
accounted for.  It is seen that the structure factor universally
suppresses the contribution from small-$q$ scattering. The suppression
sets in for larger $q$ at larger values of $\Gamma$. The large-$q$
asymptotics is independent of $\Gamma$ as $S(q)\to 1.$ The major
difference arises for intermediate values of $q$ where the structure
factor oscillates and the amplitude of oscillations increases with the
value of $\Gamma$ parameter. In addition to structure factor the
scattering matrix is folded with the nuclear formfactor, which
accounts for the finite-size of individual nucleus. Its effect on the
scattering matrix is small and is discussed in some detail in Appendix
\ref{app:A}.

\subsection{Polarization tensor }

The screening of longitudinal and transverse interactions is
determined by the corresponding components of the photon polarization
tensor. The expression \eqref{relax_final} is exact with respect to
the form of the polarization tensor. We will use an approximation to
Eq.~\eqref{relax_final} derived within the HTL effective field theory
of QED~\cite{1990NuPhB.337..569B,1990NuPhB.339..310B} in Appendix
\ref{app:B}; see also the related work on astrophysical relativistic,
dense gases of
Refs.~\cite{1992APh.....1..133A,1993APh.....1..289A,1994APh.....2..175A}.
Our computations, outlined in detail in Appendix \ref{app:B}, are
carried out at nonzero temperature and density and include the mass
of leptons (electrons and positrons); formally, we require the
four-momentum of the photon to be small compared with the
four-momentum of the fermions in the loop. For the longitudinal and
transversal components of the polarization tensor we find
\bea\label{eq:pol_final_l}
\Pi_L(q,\omega)
&=&\left(1-x^2\right) \int_0^\infty\!\!\! dp 
{\cal F} (\varepsilon) 
\left[1-\frac{x}{2v}\log\frac{x+v}{x-v}\right],\\
\label{eq:pol_final_t}
\Pi_T(q,\omega) &=&\frac{1}{2}\int_0^\infty\!\!\! dp 
{\cal F} (\varepsilon)  \left[x^2
+\left(v^2-x^2\right)\frac{x}{2v}
\log\frac{x+v}{x-v}\right],\nonumber\\
\eea
where $x=\omega/q$ and $v = \partial \ep/\partial p = p/\ep$ is the particle
velocity and 
\bea
{\cal F} (\varepsilon)  \equiv -\frac{4e^2}{\pi} p^2\left[
\frac{\partial f^+(\varepsilon)}{\partial\varepsilon}+
\frac{\partial f^-(\varepsilon)}{\partial \varepsilon}\right].
\eea
In the degenerate or ultrarelativistic limits the velocity  has a
constant value $\bar v$
and Eqs.~\eqref{eq:pol_final_l} and \eqref{eq:pol_final_t}  can be written as 
\bea\label{eq:pol_deg_ultra_l}
\Pi_L &=&q_D^2\left(1-x^2\right)  
\left[1-\frac{x}{2\bar{v}}\log\frac{x+\bar{v}}{x-\bar{v}}\right],\\
\label{eq:pol_deg_ultra_t} 
\Pi_T &=&\frac{1}{2} q_D^2\left[x^2
+\left(\bar{v}^2-x^2\right)\frac{x}{2\bar{v}}
\log\frac{x+\bar{v}}{x-\bar{v}}\right],
\eea
where $\bar{v}=v_F$ in the degenerate and $\bar{v}=1$
in the ultrarelativistic limits , respectively,  and the Debye 
wave number  $q_D$ is given by radial part of the phase-space integral 
\bea\label{eq:Debye}
q_D^{2} =\int_0^{\infty}\!\! dp ~{\cal F} (\varepsilon) .
\eea
Dropping the contribution of antiparticles we find in the limiting
cases  of highly degenerate and non-degenerate matter
\bea\label{eq:Debye_limits} q_D^{2}\simeq 4e^2
\left\{\begin{array}{ll} p_F\, \varepsilon_F/\pi,
         &\quad T\ll T_F,\\
         \pi\,  n_e\, /T , &\quad  T\gg T_F,\end{array}\right.  
\eea 
where in the last line we introduced the electron number density ($f^0 \equiv f^+$)
\bea\label{norm} 
n_e = 2\int\!\!\frac{d\vecp}{(2\pi)^3}f^0(\varepsilon).
\eea
Equations \eqref{eq:pol_deg_ultra_l} and \eqref{eq:pol_deg_ultra_t}
coincide with Eqs.~(8) and (9) of Ref.~\cite{1993APh.....1..289A}, if
we take into account the first line of Eq.~\eqref{eq:Debye_limits} and
substitute $e^2\to e^2/4\pi$ in our equations.  Note that Eqs.
\eqref{eq:pol_deg_ultra_l} and \eqref{eq:pol_deg_ultra_t} can be also
applied in the general case, if $\bar{v}$ is defined as the
characteristic velocity of electrons.
 
At temperatures of interest it is more economical to use low  
$x\ll 1$ expansions for the polarization tensor
 (see Appendix \ref{app:A}); we keep 
the next-to-leading in $x$ terms and find 
\bea \label{eq:Pi_lt}
\Pi'_L (q,\omega) = q_D^2\chi_l, \qquad \Pi_T (q,\omega) = q_D^2\chi_t, 
\eea 
where the susceptibilities to order
$O(x^2)$ are given by
\bea 
\label{eq:chi_l}
&&{\rm Re}\chi_l (q,\omega) = 1-\frac{x^2}{\bar{v}^2},
\quad {\rm Im}\chi_l (q,\omega) =-\frac{\pi x}{2\bar{v}},\\
\label{eq:chi_t}
&&{\rm Re}\chi_t (q,\omega) = x^2,  \qquad {\rm Im}\chi_t
(q,\omega) = \frac{\pi}{4}x\bar{v}.  
\eea 
Because the terms containing $\bar{v}$ are small as well
as electrons are ultrarelativistic  in the most of the regime of interest 
we approximate $\bar{v} =  1$ in our numerical calculations.

\section{Results}
\label{sec:results}

Numerically the electrical conductivity is evaluated using the
relaxation time Eq.~\eqref{relax_final} in the most general case with
the ion structure factor given in Fig.~\ref{fig:Sq} and polarization
tensor given by Eqs.~\eqref{eq:Pi_lt}--\eqref{eq:chi_t}.  With this
relaxation time we evaluate the components of the conductivity tensor
using Eq.~\eqref{eq:sigma2}.  We recall that for large magnetic fields
the tensor structure of the conductivity  is important, while in
the limit of negligible fields only the single quantity
$\sigma = \sigma_0+\sigma_2$ is relevant, see Eq.~\eqref{eq:sigma}.

\subsection{Low-field limit}

We start our analysis of the numerical results with the density,
temperature, and composition dependence of conductivity $\sigma$ in low
magnetic fields given by Eq.~\eqref{eq:sigma}. We relegate to the next
section the discussion of the $\sigma_0$ and $\sigma_1$ components, 
which is straightforward after we clarified the basic features of $\sigma$.  We
will also first study the cases of $\isotope[56]{Fe}$ and
$\isotope[12]{C}$ and later on consider density-dependent composition
of crustal matter in Sec.~\ref{sec:composition}.
\begin{figure}[!] 
\begin{center}
\includegraphics[width=\linewidth,keepaspectratio]{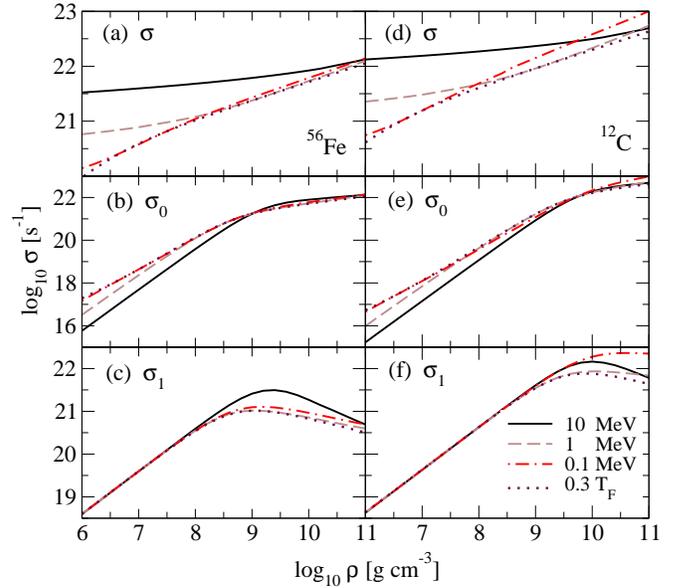}
\caption{ Dependence of three components of the electrical
  conductivity tensor on density for various values of temperature
  indicated in the plot and $B_{12}=1$ ($B_{12}\equiv
  B/10^{12}$~G).
  (a)--(c) show the conductivities for $\isotope[56]{Fe}$,
   (d)--(f) the same for $\isotope[12]{C}$. }
\label{sigma_dens_Fe_C}
\end{center}
\end{figure}
The upper panels of Fig.~\ref{sigma_dens_Fe_C} show $\sigma$ as a
function of density for various temperatures and magnetic field
$B_{12}=1$; here and below we use the units $B_{12}= B/10^{12}$~G to
characterize the magnetic field.  The temperature values range from
the non-degenerate regime ($T=10$ MeV) to the degenerate regime
($T=0.1$ MeV) where the case $T=1$ MeV is representative for
transition from non-degenerate to degenerate regime, which occurs at
around log$_{10}~\rho \simeq 8$ g cm$^{-3}$ for both $\isotope[56]{Fe}$ and
$\isotope[12]{C}$ nuclei (see Fig.~\ref{fig:PhaseDiagram}).  In each
case $\sigma$ shows a power-law dependence on density
$\sigma \propto \rho^{\alpha}$; in the degenerate regime
$\alpha \simeq 0.4$ for $\isotope[56]{Fe}$ and $\alpha \simeq 0.45$
for $\isotope[12]{C}$.  In the non-degenerate regime the increase is
less steep with $\alpha\simeq 0.1$ for both $\isotope[56]{Fe}$ and
$\isotope[12]{C}$.

This behavior can be traced back to the different density and
temperature dependence of the relaxation time in these regimes.  The
conductivity depends in both regimes on the ratio
$\tau(\varepsilon)/\varepsilon$ and is proportional to $n_e$, see 
Eqs.~\eqref{sigmas_fermi} and \eqref{sigmas_non_deg}.  For any fixed
temperature and density, the ratio $\tau(\varepsilon)/\varepsilon$
scales approximately as $\varepsilon$.  In the degenerate regime
$\varepsilon$ is the Fermi energy, therefore,
$\varepsilon\propto \rho^{1/3}$, while in the non-degenerate regime
$\bar \varepsilon \propto T$ independent of density. Apart from these
differences, $\tau^{-1}\propto n_i$ which guarantees that the
relaxation time decreases with density in both cases. These factors
combined lead to slower increase of conductivity in the non-degenerate
regime as compared to the degenerate one. Note that $\omega_c\tau$
scales as $\tau/\varepsilon$ in both cases (see Appendix~\ref{app:A} for
further details).

The main difference between the values of $\sigma$ for different
nuclei characterized by their mass number $A$ and charge $Z$ is due to
the scaling $\tau^{-1}\sim Z^2n_i\sim Z^2/A\sim Z$; for not very heavy
elements $Z/A\simeq 0.5$, see Eq.~\eqref{relax_final}. Therefore, we
find for the ratio of conductivities of $\isotope[12]{C}$ to
$\isotope[56]{Fe}$:
$\sigma_C/\sigma_{Fe}\simeq\tau_C/\tau_{Fe}\simeq Z_{Fe}/Z_C\simeq
4.3$,
which is consistent with the results shown in
Fig.~\ref{sigma_dens_Fe_C}.

\begin{figure}[!] 
\begin{center}
\includegraphics[width=\linewidth,keepaspectratio]{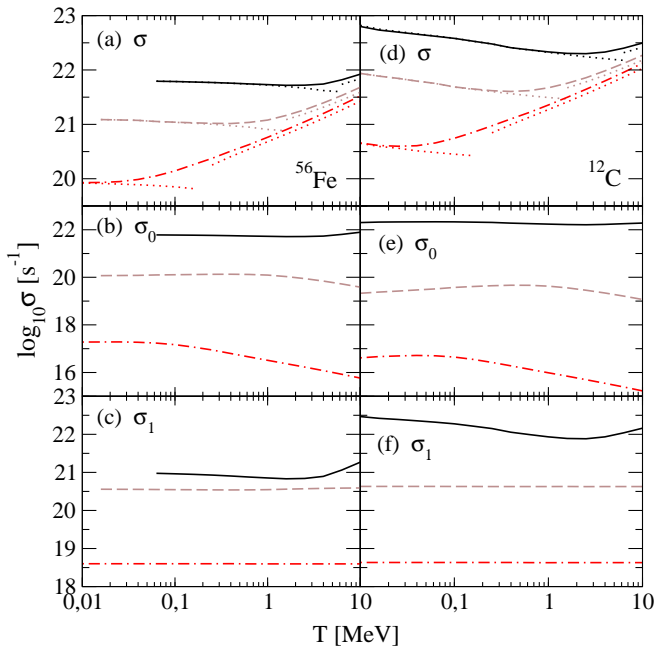}
\caption{ The temperature dependence of three components of the
  electrical conductivity tensor for $B_{12}=1$ and three values of
  density -- $\log_{10}\rho=10$ (solid lines), $\log_{10}\rho=8$
  (dashed lines), $\log_{10}\rho=6$ (dash-dotted lines) in units [g
  cm$^{-3}$].  (a)--(c) show the conductivities for
  $\isotope[56]{Fe}$, (d)--(f) the same for $\isotope[12]{C}$.
  The dots represent the results obtained with
  Eqs.~\eqref{sigmas_fermi} and \eqref{sigmas_non_deg}. }
\label{sigma_temp_Fe_C}
\end{center}
\end{figure}
\begin{figure}[!] 
\begin{center}
\includegraphics[width=7.0cm,keepaspectratio]{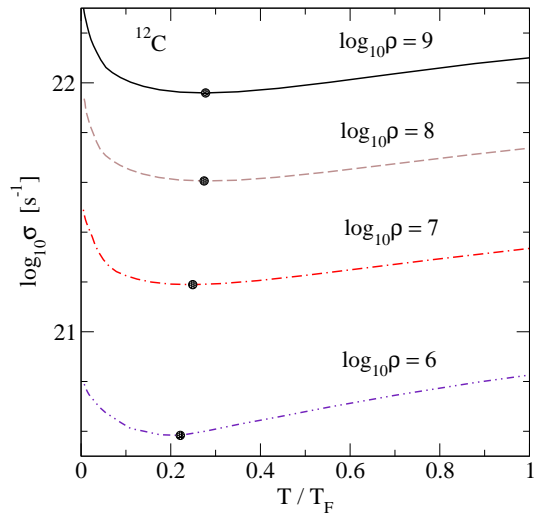}
\caption{ Dependence of conductivity on the scaled temperature for
  various densities. The minimum of the conductivity occurs roughly at
  $T\simeq T^*$.  }
\label{sigma_min}
\end{center}
\end{figure}
Let us turn to the temperature dependence of the conductivity.  The
most prominent effect seen in the temperature dependence of $\sigma$,
shown in Fig.~\ref{sigma_temp_Fe_C}, is the existence of a minimum as
a function of temperature.  The dotted lines in the low-temperature
regime correspond to the formula \eqref{sigmas_fermi} and extend to
the point where $T\simeq T_F$.  We see that the Drude formula works
very well for $T\le 0.1T_F$ and we find a good agreement between our
conductivities and those in Refs.~\cite{1984MNRAS.209..511N} and
\cite{ 1999A&A...346..345P}.  The dotted lines in the high-temperature
regime correspond to the formula \eqref{sigmas_non_deg}.  As we see
from the plots, Eq.~\eqref{sigmas_non_deg} gives the correct
qualitative behavior of the electrical conductivity at high
temperatures, but quantitatively underestimates it by about 20\%.  The
minimum in $\sigma$ arises at about the transition from the degenerate
to nondegenerate regime and is identified empirically with
$T^*\equiv 0.3T_F$. (This approximately corresponds to the requirement
that the Fermi energy becomes equal to the thermal energy of a
nondegenerate gas).  We show in Fig.~\ref{sigma_min} the dependence of
$\sigma$ on appropriately scaled temperature for a number of densities
for $\isotope[12]{C}$ (the results are similar also in the case of
$\isotope[56]{Fe}$).  We also show the density dependence of the
conductivity at the minimum in Fig.~\ref{sigma_dens_Fe_C}.  The
conductivity decreases with temperature at low temperatures, when the
electrons are degenerate. This decrease arises solely from the
temperature dependence of the correlation function $S(q)$.  In the
case $S(q) = 1$ the relaxation time is nearly temperature independent,
whereas in full calculation it decreases with temperature, as expected
(see Appendix~\ref{app:A} for numerical illustrations). Indeed, as
seen from Fig.~\ref{fig:Sq}, with increasing temperature and,
consequently, decreasing $\Gamma$ small momentum transfer scattering
becomes more important which increases the effective cross-section.

In the nondegenerate regime the temperature dependence of $\tau$
changes, because $\bar{\tau}\propto \bar{\varepsilon}^2$, therefore
conductivity $\sigma \propto \bar{\tau} /\bar{\varepsilon} \propto T$,
as suggested in Ref.~\cite{1976ApJ...206..218F} (the exact calculations give 
$\sigma\propto T^\beta$ with $\beta\simeq 0.7--0.8$).

In the degenerate regime the temperature dependence of $\sigma$ (or
$\tau$) is stronger for lighter elements, because for the given
density and temperature the parameter $\Gamma$ is smaller for lighter
elements ($\Gamma\sim Z^2/A^{1/3}$, $\Gamma_{Fe}/\Gamma_C\simeq 11$),
and the $S(q;\Gamma)$ varies faster for small values of $\Gamma$, 
see Fig.~\ref{fig:Sq}.

\subsection{Strong fields}
For strong magnetic fields the tensor structure of the conductivity
becomes important and we need to discuss the remaining components of
this tensor given by Eqs. (\ref{sigmas_fermi}) and
(\ref{sigmas_non_deg}). These components depend strongly on the value
of ``anisotropy parameter" $\omega_c\tau$. Assuming density
independent values of the magnetic fields, we find that the parameter
$\omega_c\tau$ decreases as a function of density because of the
decrease of relaxation time in any regime, see Fig.~\ref{tau_dens} of
Appendix~\ref{app:A}.  Note that in the degenerate case $\omega_c $
decreases as well because of its inverse dependence on the energy of
electrons. It is seen that at high densities $\omega_c\tau \ll 1$
(isotropic region) and $\sigma_0\simeq\sigma$.  At low densities
$\omega_c\tau\gg 1$ (strongly anisotropic region) and we have
\bea\label{sigma_0}
\sigma_0\simeq\frac{\sigma}{(\omega_c\tau)^2}\simeq\bigg(
\frac{n_ee}{B}\biggr)^2\sigma^{-1}\ll\sigma.  
\eea 
As $\omega_c\tau$ decreases with the density, 
$\sigma_0$ increases with density much
faster than $\sigma$: $\sigma_0\propto\rho^{\beta}$, $\beta\simeq 1.5$ in the degenerate and $\beta\simeq 1.9$ in the nondegenerate regime.  At low densities $\sigma_0$ is smaller than
$\sigma$ by several orders of magnitude, the exact value being 
dependent on magnetic field.
\begin{figure}[t] 
\begin{center}
\includegraphics[width=\linewidth,keepaspectratio]{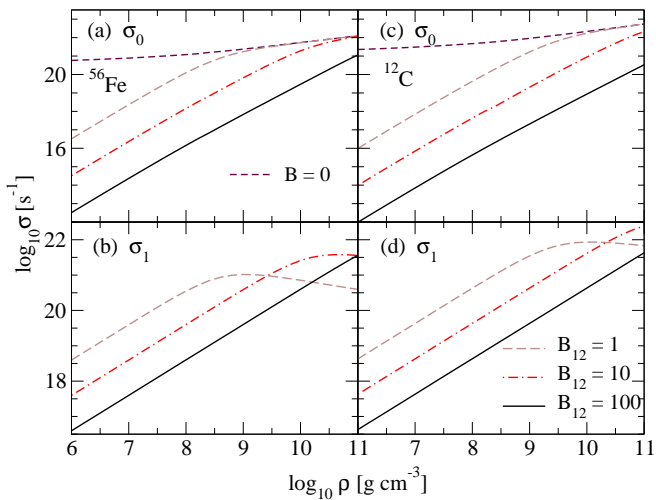}
\caption{ Dependence of $\sigma_0$ [(a) and (c)] and $\sigma_1$ [(b)
  and (d)] components of the conductivity tensor on density for
  various values of the $B$-field.  (a) and (b) show these components
  for $\isotope[56]{Fe}$, (c) and (d) the same for
  $\isotope[12]{C}$. The temperature is fixed at $T=1$ MeV.  }
\label{sigma01_dens_Fe_C}
\end{center}
\end{figure}
\begin{figure}[tbh] 
\begin{center}
\includegraphics[width=\linewidth,keepaspectratio]{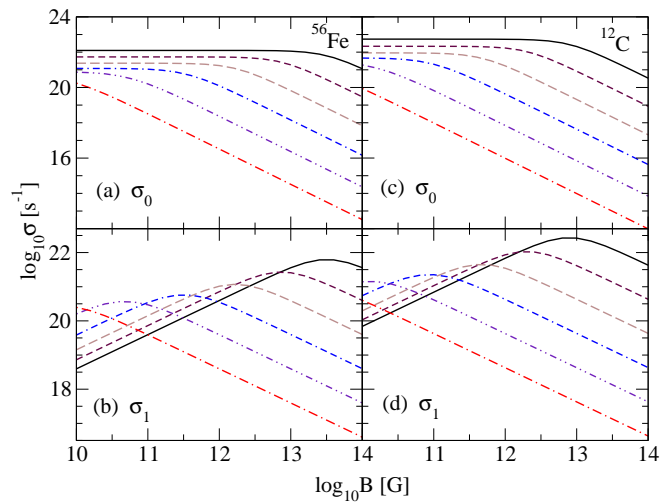}
\caption{ Dependence of $\sigma_0$ (upper panels) 
 and $\sigma_1$ (lower panels) 
components of the conductivity tensor on the magnetic field 
for $\isotope[56]{Fe}$ (left panels) 
and $\isotope[12]{C}$ (right panels) 
at six values of density -- $\log_{10}\rho=11$, 
$\log_{10}\rho=10$, $\log_{10}\rho=9$, $\log_{10}\rho=8$,
$\log_{10}\rho=7$, $\log_{10}\rho=6$ (from top to bottom). 
The temperature is fixed at $T=1$ MeV. 
}
\label{sigma01_b_Fe_C}
\end{center}
\end{figure}

We see from Eq.~(\ref{sigma_0}) that for a given density
$\sigma_0\sim \sigma^{-1}$, therefore $\sigma_0$ shows a reversed
temperature dependence at low densities. It increases in the
degenerate regime, decreases in the nondegenerate regime, see
Fig.~\ref{sigma_temp_Fe_C}, and has a maximum at temperature $T^*$. 
The reversed behavior applies also to the $Z$-dependence, \ie,
$\sigma_0\sim\tau^{-1}\sim Z$, therefore $\sigma_0$ is smaller for
lighter elements.  The curves corresponding to different temperatures
in Fig.~\ref{sigma_dens_Fe_C} intersect when $\omega_c\tau\simeq 1$ at
high density ($\rho \simeq 10^9$ g cm$^{-3}$) as a consequence of
transition from anisotropic to isotropic conduction (see also
Fig.~\ref{tau_dens}).  In addition, there are also intersections related
to the transition from degenerate (high-density) to nondegenerate
(low-density) regime, as already discussed in the case of $\sigma$.

For $\sigma_1$ component we have 
\bea\label{sigma_1_a}
\sigma_1 &\simeq& \sigma\omega_c\tau\simeq 
\frac{B}{n_ee}\sigma^2,\quad\omega_c\tau\ll1,\\
\label{sigma_1_b}
\sigma_1&\simeq&\frac{\sigma}{\omega_c\tau}\simeq
\frac{n_ee}{B}, \qquad\omega_c\tau\gg1.  
\eea 
At low densities ($\omega_c\tau\gg1$) $\sigma_1$ is proportional to
the density and depends neither on temperature, nor on the type of
nuclei, see Eq.~\eqref{sigma_1_b}. At high densities $\sigma_1$
becomes a decreasing function of density because of the additional
factor $\omega_c\tau$ in Eq.~\eqref{sigma_1_a}, the decrease being
faster at higher temperatures, see Fig.~\ref{sigma_dens_Fe_C}.  We
find the scaling $\sigma_1\propto\rho^{-\gamma}$, $\gamma\simeq 0.2$
in the degenerate and $\gamma\simeq 0.7$ in the nondegenerate regime.
As a function of density $\sigma_1$ has a maximum at
$\omega_c\tau\simeq1$, where $\sigma_0\simeq\sigma_1\simeq\sigma/2$.
In isotropic region $\sigma_1$ depends on the temperature through the
scaling $\sigma_1\sim\sigma^2$ and has a minimum at $T^*$. Because
$\sigma_1\sim Z^2$, it is larger for $\isotope[12]{C}$ as compared to
$\isotope[56]{Fe}$ by more than an order of magnitude.  As $\tau$ is
larger for light elements, the anisotropic region for these elements
is larger, and the maximum of $\sigma_1$ versus density is shifted to
higher densities and its value increases, as can be seen from
Figs. \ref{sigma_dens_Fe_C} and \ref{sigma01_dens_Fe_C}.  Note that in
both isotropic and strongly anisotropic cases $\sigma_1\ll\sigma$.
\begin{figure}[t] 
\begin{center}
\includegraphics[width=\linewidth,keepaspectratio]{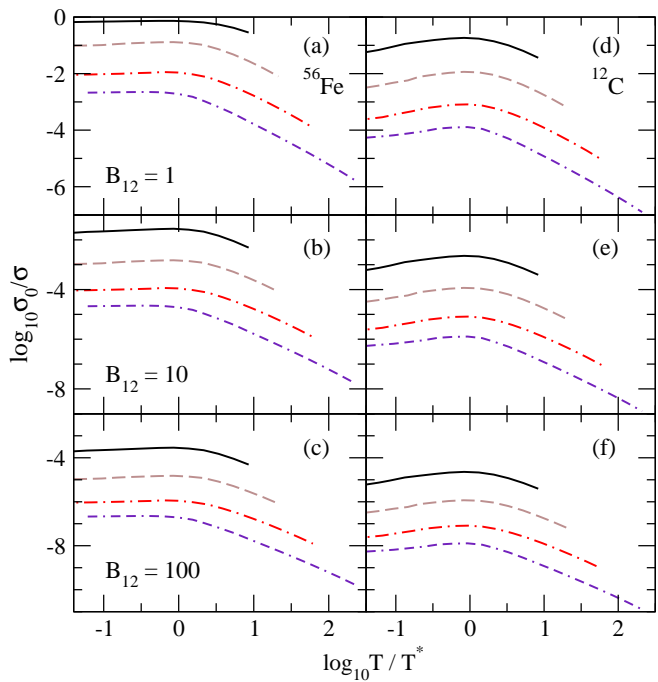}
\caption{ The ratio $\sigma_0/\sigma$ as a function of scaled
  temperature $T/T^*$ at various densities log$_{10}\rho=9$ (solid
  lines), log$_{10} \rho=8$ (dashed lines), log$_{10} \rho=7$
  (dash-dotted lines) and log$_{10} \rho=6$ (double-dash-dotted lines)
  and for three values of magnetic field $B_{12}=1$ [(a) and (d)],
  $B_{12}=10$ [(b) and (e)], and $B_{12}=100$ [(c) and (f)]. (a)--(c)
  show the ratio for $^{56}$Fe, (d)--(f) the same for $^{12}$C.}
\label{sigma0_vs_sigma}
\end{center}
\end{figure}
Figures \ref{sigma01_dens_Fe_C} and \ref{sigma01_b_Fe_C} 
show the dependence of $\sigma_0$ and
$\sigma_1$ on the magnetic field. As $\omega_c\propto B$, the density region
where the conductivity is anisotropic becomes larger with the increase of magnetic
field, and the maximum of $\sigma_1$ as a function of density 
is shifted to higher densities
(see Fig.~\ref{sigma01_dens_Fe_C}). 

For low magnetic fields $\sigma_0\simeq\sigma$
(Fig.~\ref{sigma01_b_Fe_C}).  With increasing magnetic field
$\sigma_0$ decreases and for $\omega_c\tau\gg1$ we find that
$\sigma_0\propto B^{-2}$, see Eq.~\eqref{sigma_0}.  For
$\omega_c\tau\ll1$ and $\omega_c\tau\gg1$ cases we have
$\sigma_1\propto B$ and $\sigma_1\sim B^{-1}$, respectively, see
Eqs.~\eqref{sigma_1_a} and \eqref{sigma_1_b}, therefore $\sigma_1$
should have a maximum as a function of magnetic field.  As seen from
Fig.~\ref{sigma01_b_Fe_C}, the maximum of $\sigma_1$ occurs where
$\sigma_0$ begins to drop ($\omega_c\tau\simeq 1$). This maximum
shifts to lower magnetic fields with the decrease of density and
charge $Z$.  For $B=10^{12}$ G the crust is anisotropic at
densities $\rho<10^9$ g cm$^{-3}$ for $\isotope[56]{Fe}$ and
$\rho<10^{10}$ g cm$^{-3}$ for $\isotope[12]{C}$.  For magnetic fields
$B\geq10^{13}$ G the outer crust is almost completely anisotropic.

We now turn to the study of combined effects of temperature and
magnetic field, \ie, how the anisotropy induced by the magnetic field
is affected by the temperature. To characterize the anisotropy we
consider the ratio $\sigma_0/\sigma$, which is shown in
Fig.~\ref{sigma0_vs_sigma} as a function of dimensionless ratio
$T/T^*$ for various densities and magnetic fields. We see that all
curves have a maximum at $T=T^*$ independent of density, magnetic
field, and type of nuclei. At this maximum the anisotropy is
smallest. As $\sigma_0/\sigma\propto \sigma^{-2}$, see
Eq.~\eqref{sigma_0}, it increases with the temperature in the
degenerate regime. In the nondegenerate regime $\sigma$ increases
with temperature, therefore $\sigma_0/\sigma$ decreases approximately
as $T^{-3/2}$. At very high temperatures the crust becomes strongly
anisotropic.
\begin{figure}[t] 
\begin{center}
\includegraphics[width=8cm,keepaspectratio]{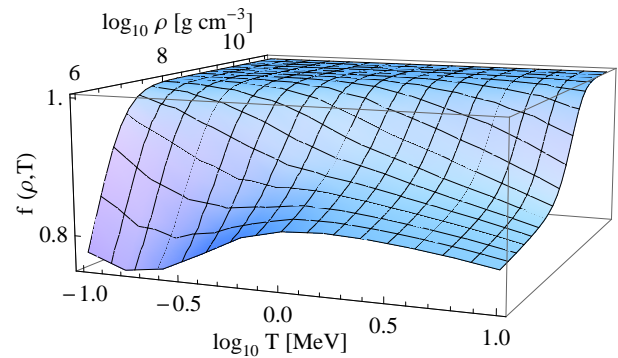}
\caption{ The function $f(\rho,T)$ for $\isotope[56]{Fe}$
at magnetic field $B_{12}=1$. }
\label{f_Fe}
\end{center}
\end{figure}

From Eqs. \eqref{sigmas_fermi} and \eqref{sigmas_non_deg} we can obtain a
simple relation between the three components of the conductivity
tensor
\bea\label{sigma_relation}
f(\rho,T)\equiv\frac{\sigma_0}{\sigma}\bigg[1+\biggl(
\frac{\sigma_1}{\sigma_0}\biggr)^2\bigg]= 1.  
\eea 
At temperatures close to the Fermi temperature 
Eqs.~\eqref{sigmas_fermi} and \eqref{sigmas_non_deg} break down, however, 
according to Fig.~\ref{f_Fe} the relation \eqref{sigma_relation} is satisfied
quite well in the whole crust. While Fig.~\ref{f_Fe} shows the case
for $^{56}$Fe, we have verified that similar results hold for $^{12}$C
and composition dependent crust and are weakly dependent on the
magnitude of the magnetic field $1\le B_{12}\le 100$.

\subsection{Density-dependent composition}
\label{sec:composition}

We now turn to the case where the composition of matter depends on the
density. We will assume that the composition does not depend strongly
on the temperature in the range of temperatures studied here ($T\le
10$ MeV) and will proceed with composition derived for $T=0$.  The
conservation of baryon number, electric charge and the condition of
$\beta$-equilibrium uniquely determines the energetically most
preferable state of matter for any given model of nuclear forces in
the density range of interest $10^6\le \rho\le \rho_{\rm drip} $,
where $\rho_{\rm drip} \simeq 4\times 10^{11}$ g cm$^{-3}$ is the
neutron drip density.

The laboratory information on nuclear masses can be used as an input
to eliminate the uncertainties related to the nuclear
Hamiltonian~\cite{2013PhRvL.110d1101W}, therefore various studies of
the composition of the crusts predict nearly identical sequences of
nuclei as a function of density.

In our calculations we adopted the nuclear sequence shown in
Fig.~\ref{fig:Phase_Diag_Comp} taken from
Ref.~\cite{2011PhRvC..83f5810P}, which predicts matter composed of
iron below the density log$_{10}\rho\le 7$ which is followed by a
sequence of nuclei with charges in the range $28\le Z\le 36$. This
composition can be compared to the initial studies of nuclear
sequences below neutron drip
density~\cite{1971ApJ...170..299B,1986bhwd.book.....S} (displayed in
Table 2.1 of Ref.~\cite{1986bhwd.book.....S}) and  a more recent study
based on improved data and theory~\cite{2006PhRvC..73c5804R}. These
deviate from the composition adopted here only marginally.
\begin{figure}[!] 
\begin{center}
\includegraphics[width=8.5cm,keepaspectratio]{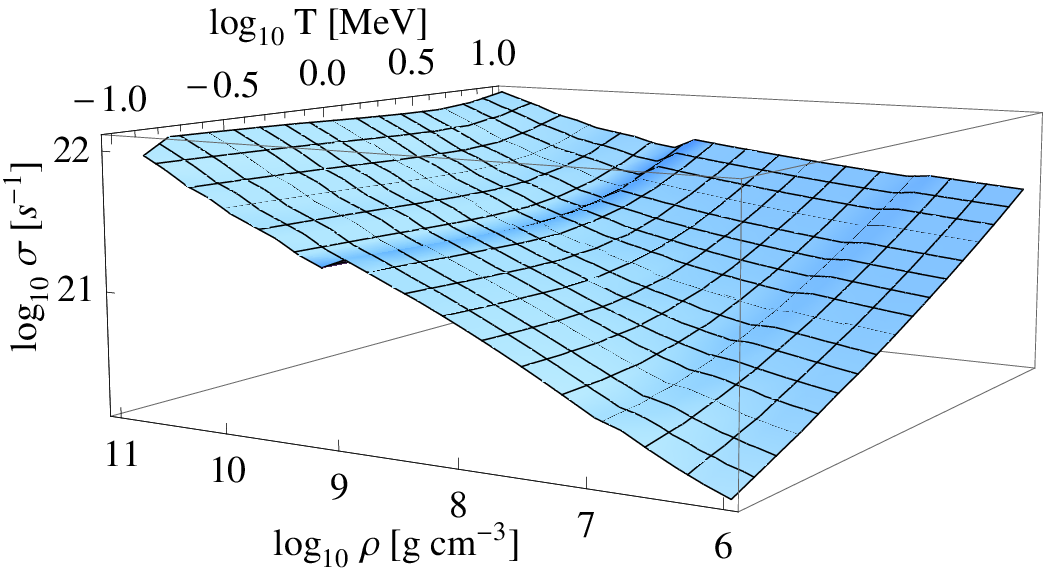}
\includegraphics[width=8.5cm,keepaspectratio]{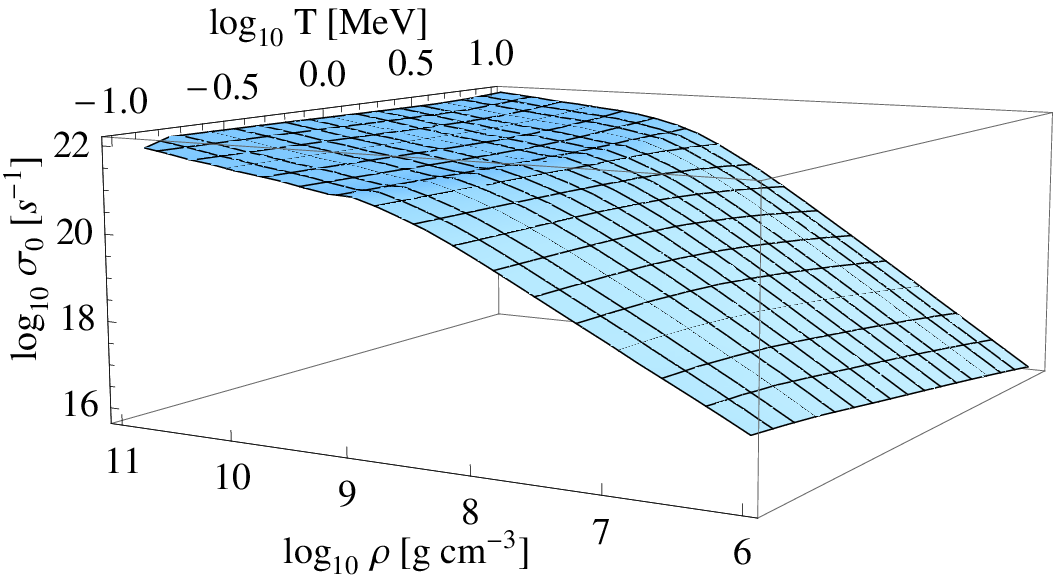}
\includegraphics[width=8.5cm,keepaspectratio]{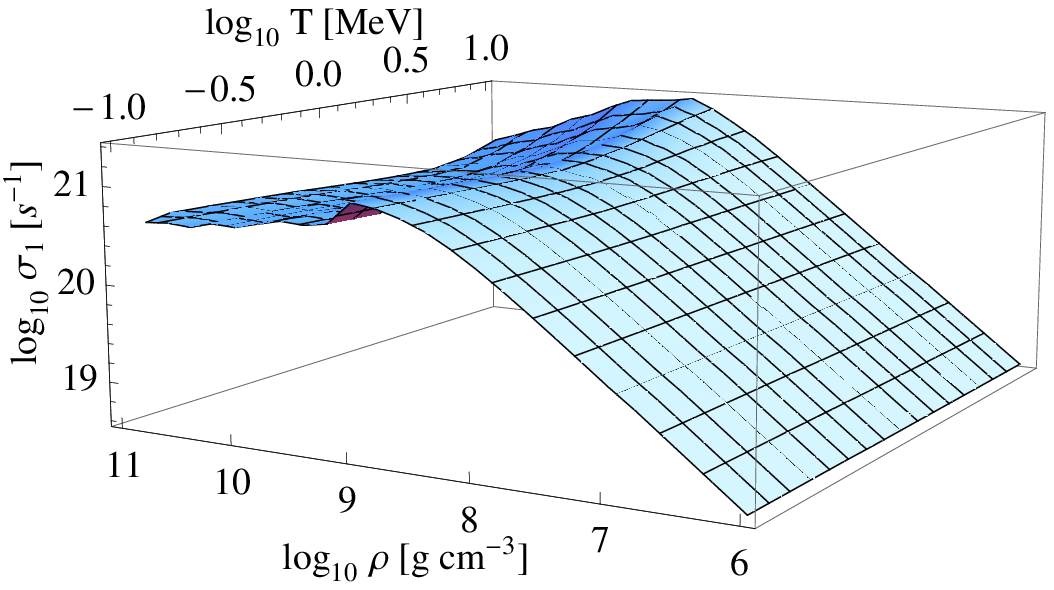}
\caption{ Dependence of three components of the electrical
  conductivity tensor on density and temperature for $B_{12}=1$ and
  composition taken from Ref.~\cite{2011PhRvC..83f5810P}.  }
\label{sigma_dens_comp}
\end{center}
\end{figure}

To assess the differences that arise from the replacement of, for
example, iron nuclei studied above by density-dependent composition
recall that for nuclei with mass number $A$ and charge $Z$ the
relaxation time scales as $\tau^{-1}\propto Z^2n_i\propto Z^2/A
$.
Because $\tau \propto \varepsilon^2$, in the degenerate regime there
will be additional density dependence in the conductivity arising from
the factor $\tau/\varepsilon \propto n_e^{1/3}$. For the conductivity
in the degenerate regime we find the scaling
\bea\label{eq:sigma_scaling_deg} \sigma\propto
\frac{n_e\tau_F}{\varepsilon_F}\propto
\left(\frac{Z}{A}\right)^{1/3}Z^{-1}.  \eea In the nondegenerate
regime \bea\label{eq:sigma_scaling_non_deg} \sigma\propto
\frac{n_e\bar\tau}{\bar\varepsilon}\propto Z^{-1}.  \eea 
To give a few numerical examples, we quote the ratio ${\cal R}$ of
conductivities of elements present in density-dependent matter
composition to that of iron: 
${\cal R}[\isotope[62]{Ni}]$ = 0.92,
${\cal R}[\isotope[64]{Ni}]$ = 0.91,
${\cal R}[\isotope[86]{Kr}]$  = 0.70,
${\cal R}[\isotope[84]{Se}]$ = 0.73,
${\cal R}[\isotope[82]{Ge}]$ = 0.77,
${\cal R}[\isotope[80]{Zn}]$  = 0.81,
${\cal R}[\isotope[80]{Ni}]$ = 0.85  
in the
degenerate regime and
${\cal R}[\isotope[62]{Ni}]$ = 
${\cal R}[\isotope[64]{Ni}]$ = 0.93,
${\cal R}[\isotope[86]{Kr}]$  = 0.72,
${\cal R}[\isotope[84]{Se}]$ = 0.76,
${\cal R}[\isotope[82]{Ge}]$ = 0.81,
${\cal R}[\isotope[80]{Zn}]$  = 0.87,
${\cal R}[\isotope[80]{Ni}]$ = 0.93 
in the nondegenerate regime.  The discrepancies between these
estimates based on the scalings
\eqref{eq:sigma_scaling_deg}, \eqref{eq:sigma_scaling_non_deg} and our
numerical results are smaller than 5\% and arise from the additional
dependence of the relaxation time on $Z$ and $A$ via the structure
factor and the Debye momentum. We conclude that the composition
dependent results differ from those found for iron by a factor
$\le 1.4$.

The three components of the conductivity tensor in the case of density-dependent composition are shown in Fig.~\ref{sigma_dens_comp} and,
according to the arguments above, show all the basic features already
discussed in the case of $\isotope[56]{Fe}$.

\subsection{Fit formulas for the electrical conductivity tensor}
\label{sec:fits}

We have performed fits to the first component of the 
conductivity tensor using the formula
\bea\label{eq:sigma_fit}
\sigma^{\rm fit}= CZ^{-1}T_F^a\bigg(\frac{T}{T_F}\bigg)^{-b}
\bigg(\frac{T}{T_F}+d\bigg)^{b+c},
\eea
where $C=1.5\times 10^{22}$ s$^{-1}$, $T$, $T_F$ are in MeV units and
$\sigma^{\rm fit}$ is in s$^{-1}$. The density dependence of
$\sigma^{\rm fit}$ arises from its dependence on the Fermi
temperature, which in general has the form
$T_F=0.511\big[\sqrt{1+(Z\rho_6/A)^{2/3}}-1\big]$. For
ultrarelativistic electrons this simplifies to
$T_F=0.511 (Z\rho_6/A)^{1/3}\simeq 0.4\rho_6^{1/3}$, where to obtain
the second relation we assumed for simplicity $Z/A \simeq
0.5$.
Substituting this into Eq.~\eqref{eq:sigma_fit} we obtain a fit
formula with explicit dependence on density
\bea\label{eq:sigma_fit_ultra}
\sigma^{\rm fit} &=& C'Z^{-1} \rho_6^{(a+b)/3}T^{-b} 
\left(T\rho_6^{-1/3}+d'\right)^{b+c},
\eea
where $C' = 0.4^{a-c} C$ and $d' = 0.4d$.

The fit parameters
$a,b,c,d$ depend on the ionic structure of the material. 
The maximal relative error of the fit formula is defined as 
$\gamma=100\vert\sigma^{\rm fit}-\sigma\vert/\sigma$. The values of
fitting coefficients in various cases are as follows:
\begin{itemize}
\item Matter composed of $\isotope[12]{C}$, $\gamma\simeq 8\%$,
\bea\label{eq:fit_constants_C}
a=0.919,\quad b=0.372,\quad c=0.813,\quad d=0.491.
\eea
\item Matter composed of $\isotope[56]{Fe}$, $\gamma\simeq 10\%$,
\bea\label{eq:fit_constants_Fe}
a=0.931,\quad b=0.149,\quad c=0.850,\quad d=0.832.
\eea
\item $\beta$-equilibrium composition, $\gamma\simeq 10\%$,
\bea\label{eq:fit_constants_comp}
a=0.935,\quad b=0.126,\quad c=0.852,\quad d=0.863.
\eea
\end{itemize}

The form of Eq.~\eqref{eq:sigma_fit} provides the correct temperature
and density dependence of the conductivity in the limiting cases of
strongly degenerate and nondegenerate electrons. For the first case
$T\ll T_F$ and $\sigma\propto T_F^{a+b}T^{-b}$. In the nondegenerate
limit $T\gg T_F$ and $\sigma\propto T_F^{a-c}T^{c}$.  As to the explicit
density dependence in these limits, one finds that
$\sigma\propto \rho^{(a+b)/3}T^{-b}$ when $T\ll T_F$ and
$\sigma\propto \rho^{(a-c)/3}T^{c}$ when $T\gg T_F$ assuming
ultrarelativistic limit. Averaging the fit
parameters we finally quote the rough scaling of the conductivity in
the limiting cases: $\sigma \propto \rho^{1/3}T^{-1/3}$ for
$\isotope[12]{C}$ and $\sigma \propto \rho^{1/3}T^{-1/7}$ in the other
cases in the degenerate regime and $\sigma \propto \rho^{1/30} T^{4/5}$ in the
nondegenerate regime.

For the other two components of the conductivity tensor
the following formulas can be used
\bea\label{eq:sigma0_fit}
\sigma_0^{\rm fit}=\frac{\sigma'}{1+\delta^2\sigma'^2},\quad
\sigma'=\sigma^{\rm fit}\bigg(\frac{T_F}{\varepsilon_F}\bigg)^{g},\\
\label{eq:sigma1_fit}
\sigma_1^{\rm fit}=\frac{\delta\sigma''^2}{1+\delta^2\sigma''^2},\quad
\sigma''=\sigma^{\rm fit}\bigg(1+\frac{T}{T_F}\bigg)^h,
\eea
where $\delta=B(n_eec)^{-1}$ in cgs units with 
$c$ being the speed of light,  $g=0.16$ and 
\bea
 h[\isotope[12]{C}]=0.075,\quad 
h[\isotope[56]{Fe}]=0.025,\quad 
h[{\rm comp.}]=0.045,\nonumber
\eea
where the last number refers to $\beta$-equilibrium composition. 

The relative error of Eq.~\eqref{eq:sigma0_fit} 
is $\gamma\simeq 11\%$ for $\isotope[12]{C}$ and 
$\gamma\simeq 13\%$ for $\isotope[56]{Fe}$ and
$\beta$-equilibrium composition.
The relative error of Eq.~\eqref{eq:sigma1_fit} 
is $\gamma\simeq 12\%$ for $\isotope[12]{C}$ and 
$\gamma\simeq 15\%$ for $\isotope[56]{Fe}$ and
$\beta$-equilibrium composition at temperatures $T>0.15$ MeV.

\section{Conclusions}
\label{sec:conclusions}

Motivated by recent advances in numerical simulations of astrophysical
phenomena such as mergers of neutron star binaries within the
resistive MHD framework we have computed here the conductivity of warm
matter ($10^{9}\le T\le 10^{11}$K) at densities corresponding to the
outer crusts of neutron stars and interiors of white dwarfs. Our
results apply to arbitrary temperatures above the solidification
temperature of matter and cover the transition from the degenerate to
the nondegenerate regimes. In this liquid plasma regime the
conductivity is dominated by the electrons which scatter off the
correlated nuclei via screened electromagnetic force. The correlations in the
plasma in the liquid state are included in terms of ion structure
function extracted from the data on Monte Carlo simulations of
one-component plasma (OCP). A key feature of our computation is the
inclusion of the dynamical screening of photon exchange and inelastic
processes, which we show to be small in the temperature-density regime
considered. The use of OCP structure factor implies that our
results should be applied with caution in the case where matter is
composed of mixture of nuclei, in which case the interspecies
correlations are not accounted for. We have implemented the HTL QED
polarization susceptibilities in the low-frequency limit combined with
nonzero-temperature screening Debye length, which should be a good
approximation where the inelastic processes are suppressed by the
large mass of nuclei.  A further simplifying approximation that went
into our formalism, which is well justified by the MHD regime of
astrophysical studies, is the assumption of weakly non-equilibrium
state of the plasma. This allowed us to express the solution of the
Boltzmann kinetic equation in relaxation time approximation.

We find that the conductivity as a function of the temperature shows a
minimum around $0.3T_F$ almost independent of the density and
composition of matter, which arises as a result of the transition from
the degenerate regime ($T\ll T_F$) to the nondegenerate regime ($T\gg
T_F$). Thus, the conductivity decreases with increasing temperature
in the degenerate regime up to the point $0.3T_F$; further increase in
the temperature leads to a power-law increase in the conductivity as
the system enters the nondegenerate regime.  We further find that at
fixed temperature the conductivity always increases with density, but
the slope of the increase is weaker in the nondegenerate regime.

We have further extracted the components of the conductivity tensor in
the entire density and temperature range for nonquantizing fields
$10^{10}\le B\le 10^{14}$~G. Because the product of relaxation time
and cyclotron frequency is a decreasing function of density in the
complete temperature range, low-density matter features anisotropic
conductivity at lower magnetic fields. For example, the component of
the conductivity transverse to the field $\sigma_0\to \sigma$ in the
high density limit, but is substantially suppressed at low
densities. This underlines the importance of proper inclusion of
anisotropy of conductivity in astrophysical studies of dilute
magnetized matter even at relatively low magnetic fields.

 Our results can be implemented in numerical studies in terms of fit
 formulas \eqref{eq:sigma_fit}--\eqref{eq:sigma1_fit}. An alternative
 is to use plain text tables of conductivities, see Supplemental
 Material \cite{SupMat}.

 Finally, our results show that the conductivity depends weakly on the
 composition of matter.  For example, the conductivity of matter
 composed of heavy elements with $26\le Z\le 36$ in
 $\beta$-equilibrium with electrons differs from the conductivity of
 matter composed of $^{56}$Fe at the same density and temperature by a
 factor $\le 1.4$. It would be interesting, however, to study the
 conductivity of warm multicomponent matter which is composed of
 nuclei in statistical equilibrium, in which case composition may
 become an important factor.

\section*{Acknowledgements}

We thank L. Rezzolla (L.R.), D. H. Rischke and J. Schaffner-Bielich for
useful discussions, L.R. for drawing our attention to this problem, 
 and M.~N.~Tamashiro and Y.~Levin for useful
communications. A. H. acknowledges support from the HGS-HIRe graduate
program at Frankfurt University. A. S. is supported by the Deutsche
Forschungsgemeinschaft (Grant No. SE 1836/3-1). We acknowledge the
support by NewCompStar COST Action MP1304.

\appendix
\section{Evaluating the relaxation time}
\label{app:A}

The purpose of this appendix is to give the details of the transition
from the relaxation time \eqref{eq:t_relax} to
Eq.~\eqref{relax_final}. We start by defining several angles by the
relations $ \vecp \cdot \vecq = p q \cos \alpha,$
$\vecp_t\cdot \vecp'_t = p_t p'_t \cos\phi $ and
$\vecq\cdot \vecp' = q p' \cos \vartheta$, where $\vecp_t$, $\vecp'_t$
are the components of $\vecp$, $\vecp'$ transversal to $\vecq$.
Writing the  second $\delta$-function in  Eq.~\eqref{eq:t_relax} as 
$\delta(\varepsilon_2-\varepsilon_4+\omega)
=({M}/{p'q})\delta(\cos\vartheta-x_0)$, where
$x_0=(2\omega M-q^2)/{2p'q}$, we find 
\bea\label{relax3}
\tau^{-1}(\varepsilon) &=&
\frac{(2\pi)^{-5} M}{p} \int d\omega 
d\bm q  \cos\alpha \delta(\varepsilon-\varepsilon_3-\omega) 
\frac{1-f^0_3}{1-f^0}\nonumber\\
&\times&
\int_0^\infty   dp'  p' g(p') S(q)F^2(q) I_\Omega,
\eea
where 
\bea\label{angle_integral}
I_{\Omega}&=&\int d (\cos\vartheta) d\phi
 \frac{Z^2e^{*4}}{2\varepsilon\varepsilon_3}\Biggl[
 \frac{2\ep^2-\ep\omega -p q \cos \alpha }{|q^2+\Pi'_L|^2} 
\nonumber\\
&+&\frac{p'^{2}\sin^2\vartheta[ 2(p\cos\phi)^2\sin^2\alpha 
+\left( -\ep  \omega +q p \cos \alpha\right)]
}{M^2|q^2-\omega^2+\Pi_T|^2}
\nonumber\\
 &-& \frac{2}{M}
\frac{(2\ep-\omega) (p p'\sin\alpha\sin\vartheta\cos\phi
 )}{{\rm Re} (q^2+\Pi'_L)(q^2-\omega^2+\Pi_T^*)}
\Biggr] \delta(\cos\vartheta-x_0),\nonumber\\
\eea
and we substituted the expression for the matrix element \eqref{probabiliy3}.
After integration over the angle $\phi$ we obtain
\bea\label{angle_integral}
I_{\Omega} 
&=&\pi \frac{Z^2e^{*4} }{\ep(\ep-\omega)} 
\Biggl[
 \frac{2\ep^2-\ep\omega -p q \cos \alpha }{|q^2+\Pi'_L|^2}
\nonumber\\
&+&p'^2(1-x_0^2)\frac{p^2 \sin^2\alpha +
  q p \cos\alpha -\ep  \omega 
}{M^2|q^2-\omega^2+\Pi_T|^2}\Biggr] \theta(1-\vert x_0\vert ).
\nonumber\\
\eea
The step-function $\theta$ defines the minimum value
$
p'_{\rm min } = \vert {2\omega  M-q^2}\vert /{2q} 
$
for the integration over this variable. 
We substitute Eq.~\eqref{angle_integral} in 
Eq.~\eqref{relax3}, implement the integration bound on $p'$ and find 
\bea\label{relax4}
\tau^{-1}(\varepsilon)  &=&
\frac{(2\pi)^{-5}\pi M}{p}  \frac{Z^2e^{*4}  }{\ep} 
\nonumber\\
&\times&
\int\!\! d\omega 
d\bm q\frac{ S(q)F^2(q)\cos\alpha}{(\ep-\omega)} 
\delta(\varepsilon-\varepsilon_3-\omega)
\frac{1-f^0_3}{1-f^0}
\nonumber\\
&\times&\int_{p'_{\rm min}}^{\infty}  
 \!\!\! \!\! dp'  p' g(p')  \Biggl[
 \frac{2\ep^2-\ep\omega -p q \cos \alpha }{|q^2+\Pi'_L|^2}
\nonumber\\
&+&p'^2(1-x_0^2)\frac{p^2 \sin^2\alpha +
  q p \cos\alpha -\ep  \omega 
}{M^2|q^2-\omega^2+\Pi_T|^2}\Biggr] .
\eea
The remaining $\delta$ function is written as
$
\delta(\varepsilon-\varepsilon_3-\omega)=
[(\varepsilon-\omega)/{pq}]\delta(\cos\alpha-y_0)
\vartheta(\varepsilon-\omega)$ where 
$y_0=(q^2-\omega^2+2\varepsilon\omega)/{2pq}.$
In the next step we integrate over $p'$ to obtain 
\bea\label{relax7}
\tau^{-1}(\varepsilon) 
&=&\frac{ M^2}{8(2\pi)^{3}  p^3\beta}  
\frac{Z^2e^{*4} }{\varepsilon} 
\int_{-\infty}^{\varepsilon} 
d\omega 
\frac{1-f(\varepsilon-\omega)}{1-f^0(\varepsilon)}
\nonumber\\
&\times& \int_0^\infty d q    ~S(q)F^2(q)~  ~g(p'_{\rm min}) 
(q^2-\omega^2+2\varepsilon\omega)
\nonumber\\
&\times&
\theta(1-\vert y_0\vert) \Biggl[ \frac{(2\ep-\omega)^2-q^2}{|q^2+\Pi'_L|^2}
+\frac{1}{\beta M q^2}  \nonumber\\
&\times&
\frac{ 
(q^2 -\omega^2 ) [ (2\varepsilon - \omega)^2 + q^2  ]
  -4q^2m^2 }{|q^2-\omega^2+\Pi_T|^2}\Biggr].
\eea
Finally the $\theta$-function puts 
some limitations on the integration region over $q$, specifically 
$q_-\le q\le q_+$,  where 
$q_{\pm}  = \vert \pm   \sqrt{\ep^2-m^2} + \sqrt{(\ep-\omega)^2-m^2}\vert$.
Note also that to have a real $q$ we need 
$\omega \le \ep-m$. Implementing these limits we 
obtain 
\bea\label{eq:relax_time4} 
\tau^{-1}(\varepsilon) &=&\frac{
  M^2T}{8(2\pi)^{3} p^3} \frac{Z^2e^{*4}}{\varepsilon}
\int_{-\infty}^{\varepsilon-m} d\omega
\frac{1-f^0(\varepsilon-\omega)}{1-f^0(\varepsilon)}
\nonumber\\
&\times&
\int_{q_-}^{q_+} d q
~S(q)~ F^2(q)~g(p'_{\rm min}) (q^2-\omega^2+2\varepsilon\omega)
\nonumber\\
&\times& \Biggl[ \frac{(2\ep-\omega)^2-q^2}{|q^2+\Pi'_L|^2} 
+\frac{T}{M
  q^2}
\nonumber\\
&\times& \frac{ (q^2 -\omega^2 ) [ (2\varepsilon - \omega)^2 + q^2 ]
  -4q^2m^2 }{|q^2-\omega^2+\Pi_T|^2}\Biggr] .
\eea 
Finally we write Eq.~\eqref{eq:maxwell} as
\bea\label{maxwell2}
g({p'_{\rm min}})
=
n_i\bigg(\frac{2\pi}{MT}\bigg)^{3/2}
e^{-x^2/2\theta^2}e^{\omega/2T}e^{-q^2/8MT},
\eea
where $x={\omega}/{q}$ and $\theta =\sqrt{T/M}$ and substitute 
in Eq.~\eqref{eq:relax_time4} to  obtain Eq.~\eqref{relax_final} 
of the main text.
\begin{figure}[!] 
\begin{center}
\includegraphics[width=7cm,keepaspectratio]{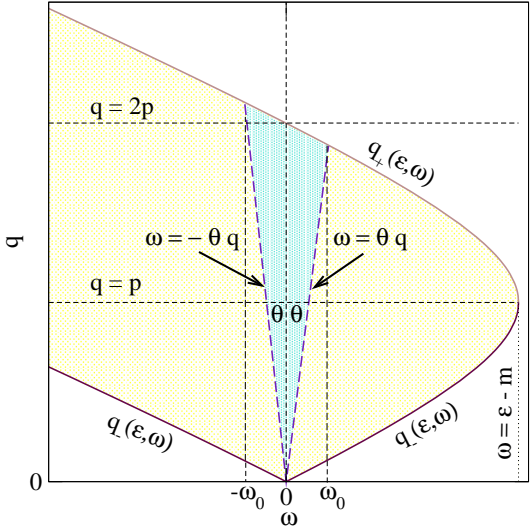}
\caption{ The integration region in Eq.~\eqref{relax_final} in the
  plane $(\omega,q)$ (light shaded area) bounded by the functions
  $q_{\pm} (\ep)$.  The major contribution to the integral comes from
  the triangle bound by the lines $\omega = \pm\theta q$ with the
  narrow opening angle $2\theta$ (dark shaded area).  }
\label{fig:integral_limits}
\end{center}
\end{figure}

In our numerical calculations the temperature varies in the range
$0.1\le T\le 10$ MeV and the masses of the nuclei lie in the range
from $10^4$ MeV to $10^5$ MeV. Therefore the parameter $\theta$
changes in the range $10^{-3}<\theta<3\times 10^{-2}\ll 1$. As a result
one can expect that the dynamical part of the scattering amplitude
should be suppressed compared with the static part by several orders
of magnitude. Numerical calculations show that the contribution of the
dynamical part is smaller than 0.15\% for $\isotope[12]{C}$ and 0.04\%
for $\isotope[56]{Fe}$ and have the order of $\theta^2$, as expected.
Due to the exponent $e^{-x^2/2\theta^2}$ of the expression
(\ref{maxwell2}) only small values of $x$ ($|x|<\theta$) contribute
significantly to the integral (\ref{relax_final}).  Therefore, the
effective phase volume of the double integration in
Eq.(\ref{relax_final}) reduces to the triangle limited by the lines
\bea \omega=\pm\theta q,\quad
q_+(\varepsilon,\omega)= 2p-\omega v^{-1}\approx 2p.  
\eea 
as illustrated in Fig.~\ref{fig:integral_limits}.  It seen from this
figure that the effective width of $\omega$ variable is given
by $\omega_0=2p\theta$.

In Eq.~\eqref{relax_final} we can take the limit where
the ionic component of the plasma is 
at zero temperature (the temperature of the electronic
component is arbitrary so far). Then, because 
\bea\label{delta_exp}
\lim_{\theta\to 0}\frac{1}{\theta\sqrt{2\pi}}{\rm e}^{-x^2/2\theta^2}
 = \delta(x),
\eea
we obtain 
\bea\label{eq:relax_recoil}
\tau^{-1}(\varepsilon)
&=&\frac{\pi Z^2e^4n_i}{\varepsilon p^3}
\int_{0}^{2p}\!\!\! dqq^3S(q)F^2(q)\nonumber\\
&\times & e^{-q^2/8MT}
\frac{4\varepsilon^2 -q^2}{|q^2+\Pi'_L|^2},
\eea
where we took into account the fact that for $\omega=0$ the limits on
momentum transfer reduce to $q_-=0$ and $q_+=2p$.  Neglecting also the
nuclear recoil (which amounts to replacing the exponential factor
$e^{-q^2/8MT}$ 
by unity) we obtain the well-known expression of the relaxation time
\eqref{eq:relax_time3}. 

Note that the nuclear recoil factor $e^{-q^2/8MT}$ can be important at
very low temperatures and high densities for light nuclei, where it
can reduce the scattering amplitude significantly. For example, in the
extreme case $T=0.01$ MeV and $\log_{10} \rho = 11$ and
$\isotope[12]{C}$ we find that the relative error could be larger than
a factor of 2. Therefore, Eq.~\eqref{eq:relax_recoil} is a better
approximation than Eq.~\eqref{eq:relax_time3} in the static limit
$\omega\to 0$.  However, in the main density-temperature range we
consider the nuclear recoil and the dynamical screening introduce only
small corrections and do not change the general behavior of the
conductivity.  The deviations between Eqs.~\eqref{eq:relax_time3} and
\eqref{relax_final} are smaller than 12\% for $\isotope[12]{C}$ and
5\% for $\isotope[56]{Fe}$.

It can be shown that at densities $\rho <10^{11}$ g cm$^{-3}$ the
effect of nuclear form factor is small as well. Indeed, for the
heaviest nucleus that we consider ($\isotope[86]{Kr}$)
$r_c\simeq 0.025$ MeV$^{-1}$ and the maximal value of the parameter is
$qr_c\simeq 0.5$. For small $qr_c$ we can use the approximate formula
obtained from Eq.~(\ref{formfactor}) \bea\label{formfactor_approx}
F^2(q)\approx 1-0.2(qr_c)^2, \eea therefore the maximal correction for
$\isotope[86]{Kr}$ is 5\%, which is consistent with numerical results.
The corrections are smaller than 4\% for $\isotope[56]{Fe}$ and 2\%
for $\isotope[12]{C}$.

\begin{figure}[!] 
\begin{center}
\includegraphics[width=8.6cm,keepaspectratio]{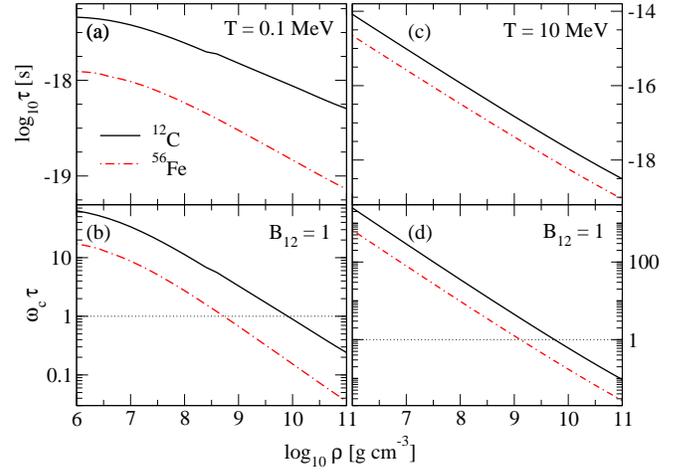}
\caption{ The relaxation time [(a) and (c)] and dimensionless
  product $\omega_c\tau$ [(b) and (d)] as functions of
  density. The temperature is fixed at $T=0.1$ MeV in (a) and
  (b) (degenerate regime) and at $T=10$ MeV in  (c) and (d)
  (nondegenerate regime).  The magnetic field is fixed at $B_{12}=1$.
}
\label{tau_dens}
\end{center}
\end{figure}
\begin{figure}[hbt] 
\begin{center}
\includegraphics[width=8.6cm,keepaspectratio]{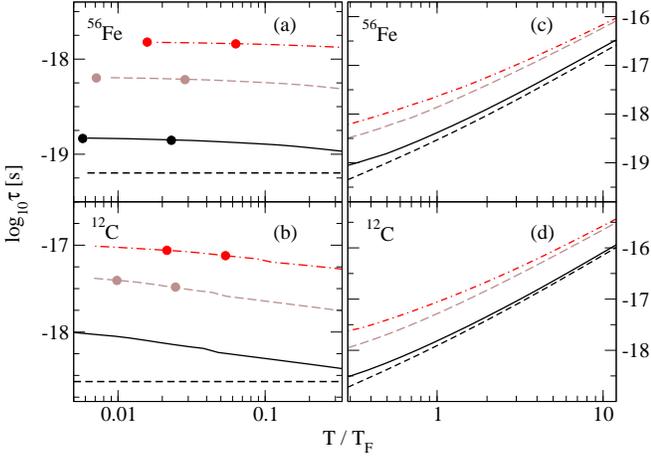}
\caption{ The relaxation time as a function on dimensionless ratio
  $T/T_F$ for three values of density: $\log_{10}\rho=10$ (solid
  lines), $\log_{10}\rho=8$ (dashed lines), $\log_{10}\rho=6$
  (dash-dotted lines) for $\isotope[56]{Fe}$ [(a) and (c)] and
  $\isotope[12]{C}$ [(b) and (d)].  (a) and (b)
  correspond to the degenerate and  (c) and (d)  to the
  nondegenerate regime.  We show the effect of setting $S(q)= 1$ for
  $\log_{10}\rho=10$ by short-dashed lines.  The open circles
  reproduce the results of Ref.~\cite{1984MNRAS.209..511N}.}
\label{tau_temp}
\end{center}
\end{figure}
Finally we provide in Figs.~\ref{tau_dens} and ~\ref{tau_temp}
numerical values of the relaxation time for a number of cases of
interest. We stress that $\tau$ is energy-dependent and is evaluated
in the degenerate case at the Fermi energy and in the nondegenerate
case at $\bar\varepsilon \simeq 3T$, which is the thermal energy of
ultrarelativistic electrons. In Fig.~\ref{tau_dens} we show the
dependence of the relaxation time and the factor $\omega_c\tau$ on
density for two values of temperature. It is seen that in the
degenerate regime ($T=0.1$ MeV) the slope of decrease in the
relaxation time is smaller than in the case of nondegenerate
regime. It is also seen that the factor $\omega_c\tau$ makes a
crossover from being much larger than unity at small densities to
being much smaller at high densities. This indicates that in
low-density matter the effects of anisotropy are much more important
than in the high-density matter. In fact, in the nondegenerate case
the low-density matter has highly anisotropic conductivity with, for
example, $\omega_c\tau\sim 10^3$ for $B_{12}=1$.

Figure~\ref{tau_temp} shows the temperature dependence of the
relaxation time for several densities. Our results agree well with
those of Nandkumar and Pethick~\cite{1984MNRAS.209..511N} in the
degenerate regime. It is seen that $\tau$ decreases as a function of
temperature in the degenerate regime for $T\le T^*$ and increases in
the nondegenerate regime $T\ge T^*$, which makes clear the existence
of the minimum at $T^*=0.3T_F$ in the conductivity. The temperature
decrease in the degenerate regime is caused almost entirely by the structure
factor $S(q)$ (see Fig.~\ref{tau_temp}).  In the nondegenerate regime
the temperature dependence of $\tau $ is dominated by the energy
increase of electrons with temperature and the role of $S(q)$ is less
important, see Fig.~\ref{tau_temp}. This is due to the fact that when
$T\ge T_F$, i.e., electrons are nondegenerate, the ionic component
forms a Boltzmann gas for composition consisting of $\isotope[56]{Fe}$
and $\isotope[12]{C}$ nuclei (see Fig.~\ref{fig:PhaseDiagram}).

\section{Polarization tensor}
\label{app:B}
In this appendix we outline the derivation of the polarization tensor
and the variant of the HTL effective field theory that underlies our
computation. Most of the HTL computation are carried out in the
ultrarelativistic (massless) limit; here we keep the particle mass
nonzero and implement HTL approximation by requiring that the
external photon four-momenta are small compared to the fermionic
four-momenta within the fermionic loop. 

The full photon propagator $D_{\mu\nu}$ can be found from the Dyson
equation
\bea\label{eq:Dyson}
[D^{-1}]_{\mu\nu}=[D_0^{-1}]_{\mu\nu}-\Pi_{\mu\nu},
\eea
where $D_0^{\mu\nu}=g^{\mu\nu}/Q^2$ is the free photon propagator with $Q^2=\omega^2-q^2$. 
$\Pi_{\mu\nu}$ is the photon polarization tensor 
 and can be decomposed into
transverse and longitudinal modes
\bea\label{eq:pol}
\Pi_{\mu\nu}=\Pi_T P^T_{\mu\nu}+\Pi_L P^L_{\mu\nu}.
\eea
We work in the plasma rest frame where the 
projectors $P^T_{\mu\nu}$ and $P^L_{\mu\nu}$
have the following components
\bea\label{eq:trans}
&&P^T_{00}=0,\quad P^T_{0i}=0,\quad P^T_{ij}=-\delta_{ij}+q_iq_j/q^2,\\
\label{eq:long}
&&P^L_{00}=-q^2/Q^2,\quad P^L_{0i}=-\omega q_i/Q^2,\quad 
P^L_{ij}=-\frac{\omega^2}{Q^2} \frac{q_iq_j}{q^2} .\nonumber\\
\eea
They satisfy  the relations 
\bea\label{eq:pol_projectors}
P^T_{\mu\alpha}P^{T\alpha\nu}=P^{T\nu}_{\mu},\quad
P^L_{\mu\alpha}P^{L\alpha\nu}=P^{L\nu}_{\mu},\quad
P^T_{\mu\alpha}P^{L\alpha\nu}=0.\nonumber\\
\eea
From Eqs.~\eqref{eq:Dyson}-\eqref{eq:pol_projectors}
it is easy to find the full photon propagator
\bea\label{eq:photon_prop}
D^{\mu\nu}=\frac{1}{Q^{2}}\bigg[g^{\mu\nu}+\frac{\Pi_T}{Q^2-\Pi_T}
P^{T\mu\nu}+\frac{\Pi_L}{Q^2-\Pi_L}P^{L\mu\nu}\bigg].\nonumber\\
\eea
Using Eq.~\eqref{eq:photon_prop} and the current conservation law
$q^\mu J_\mu=\omega J_0+q^iJ_i=0$ we can express the scattering
amplitude via two scalar functions $\Pi_L$ and $\Pi_T$:
\bea\label{eq:amplitude1}
{\cal M}=J_\mu D^{\mu\nu}J'_\nu =-\frac{J_0J'_0}{q^2+\Pi'_L}
+\frac{\bm J_{t}\bm J'_{t}}{q^2-\omega^2+\Pi_T},
\eea
where we introduced transversal currents $J_{ti}=J_j(\delta_{ij}-q_iq_j/q^2)$ and
$\Pi'_L=-\Pi_L q^2/Q^2=\Pi_L/(1-x^2)$ and $x=\omega/q$.
The one-loop diagram in the imaginary-time Matsubara formalism is given by 
\bea\label{eq:pol1}
\Pi_{\mu\nu}(\bm q,\omega_n)&=&e^{*2}\int \frac{d\bm p}{(2\pi)^3} 
T\sum\limits_{m}
\nonumber\\
&&\hspace{-1cm}\times
\Tr[\gamma_\mu S(\bm p,\omega_m)
\gamma_\nu S(\bm p-\bm q,\omega_m-\omega_n)],
\eea
where $S(\bm p, \omega_n)$ is free electron-positron propagator and the sum
goes over fermionic (odd) Matsubara frequencies $\omega_m=(2m+1)\pi T-i\mu$
[$\omega_n=2n\pi T$ is a bosonic (even)  Matsubara frequency]. The 
propagator  $S(\bm p, \omega_n)$ is given by 
\be\label{eq:freeprop} 
S(\bm p,\omega_m)=i\sum\limits_{\pm}\frac
{\Lambda^\pm_p\gamma_0}{i\omega_n-E^\pm_p},
\ee
where $\Lambda^\pm_p$ are the projection operators onto positive and
negative energy states
\be\label{eq:projectors}
\Lambda^\pm_p=\frac{\slashed p^{\pm}+m}{2E_p^\pm}\gamma_0,\quad p^{\pm}=(E_p^{\pm},\bm p)=(\pm E_p,\bm p).
\ee
\begin{widetext} 
Substitution of Eq.~(\ref{eq:freeprop}) into Eq.~(\ref{eq:pol1}) gives
\bea\label{eq:pol2}
\Pi_{\mu\nu}(\bm q,\omega_n)
=-e^{*2}\int \frac{d\bm p}{(2\pi)^3}
\sum\limits_{\pm\pm}T\sum\limits_{m}
\frac{\Tr[\gamma_\mu\Lambda^{\pm}_p
\gamma_0\gamma_\nu\Lambda^\pm_{p-q}\gamma_0]}{(i\omega_m- E^\pm_p)(i\omega_m-i\omega_n- E^\pm_{p-q})}.
\eea
The trace is evaluated using standard field theory methods after
substituting therein the projectors \eqref{eq:projectors}:
\bea\label{eq:traces}
\Tr[\gamma_\mu\Lambda^{\pm}_p
\gamma_0\gamma_\nu\Lambda^\pm_{p-q}\gamma_0]=\frac{p^{\pm}_{\mu}p'^{\pm}_{\nu}
+p^{\pm}_{\nu}p'^{\pm}_{\mu}-g_{\mu\nu}[(p^{\pm}p'^{\pm})-m^2]}{E_p^{\pm}E_{p'}^{\pm}},\qquad
\bm p'=\bm p-\bm q.
\eea
The summation over the Matsubara frequencies gives
\bea\label{eq:matsubarasums}
T\sum\limits_{m}\frac{1}
{(i\omega_m- E^\pm_p)(i\omega_m-i\omega_n- E^\pm_{p-q})}=
\frac{f^+(E_p^{\pm})-f^+(E_{p-q}^{\pm})}{E_p^{\pm}-E^{\pm}_{p-q}-i\omega_n},
\eea
where $f^\pm(E)=[e^{\beta(E\mp \mu)}+1]^{-1}$ (note that in the main text we use
$f^0$ instead of $f^+$).   Substituting 
Eqs.~\eqref{eq:traces} and \eqref{eq:matsubarasums} 
into Eq.~\eqref{eq:pol2} we obtain 
\bea\label{eq:pol3}
\Pi_{\mu\nu}(\bm q,\omega_n)=-e^{*2}\int 
\frac{d\bm p}{(2\pi)^3}
\sum\limits_{\pm\pm}\frac{p^{\pm}_{\mu}p'^{\pm}_{\nu}
+p^{\pm}_{\nu}p'^{\pm}_{\mu}-g_{\mu\nu}[(p^{\pm}p'^{\pm})-m^2]}{E_p^{\pm}E_{p'}^{\pm}}
\frac{f^+(E^{\pm}_p)-f^+(E^{\pm}_{p-q})}
{E^{\pm}_p-E^{\pm}_{p-q}-i\omega_n}.
\eea
Consider the spatial components of Eq.~\eqref{eq:pol3}
\bea\label{eq:pol_ij}
\Pi_{ij}
=-e^{*2}\int \frac{d\bm p}{(2\pi)^3}
\bigg\{\frac{p_ip'_j+p_jp'_i-
\delta_{ij}(m^2-E_pE_{p-q}
+\bm p\cdot\bm p')}{E_pE_{p-q}}\bigg[
\frac{f^+(E_p)-f^+(E_{p-q})}
{E_p-E_{p-q}-i\omega_n}+
\frac{f^+(-E_p)-f^+(-E_{p-q})}
{-E_p+E_{p-q}-i\omega_n}\bigg]\nonumber\\
-\frac{p_ip'_j+p_jp'_i-
\delta_{ij}(m^2+E_pE_{p-q}
+\bm p\cdot\bm p')}{E_pE_{p-q}}\biggl[
\frac{f^+(E_p)-f^+(-E_{p-q})}
{E_p+E_{p-q}-i\omega_n}+
\frac{f^+(-E_p)-f^+(E_{p-q})}
{-E_p-E_{p-q}-i\omega_n}\bigg]\bigg\}.\nonumber\\
\eea
The part of the polarization tensor $\propto \delta_{ij}$ gives
\bea\label{eq:pol_ij1}
\Pi^{(1)}_{ij}(\bm q,\omega_n)
&=&-e^{*2}\delta_{ij}\int \frac{d\bm p}{(2\pi)^3}
\bigg\{-
\frac{m^2-E_pE_{p-q}+\bm p\cdot\bm p'}{E_pE_{p-q}}
\bigg[\frac{f^+(E_p)-f^+(E_{p-q})}
{E_p-E_{p-q}-i\omega_n}+\frac{f^+(-E_p)-f^+(-E_{p-q})}
{-E_p+E_{p-q}-i\omega_n}\bigg]\nonumber\\
&+&\frac{m^2+E_pE_{p-q}+\bm p\cdot\bm p'}{E_pE_{p-q}}
\biggl[\frac{f^+(E_p)-f^+(-E_{p-q})}{E_p+E_{p-q}-i\omega_n}+
\frac{f^+(-E_p)-f^+(E_{p-q})}{-E_p-E_{p-q}-i\omega_n}\bigg]
\bigg\}.
\eea
In the spirit of the HTL approximation we next put $\vecq = 0$ in the
pre-factors multiplying the occupation numbers, use the relation
$f^+(-E)=1-f^-(E)$ and drop the vacuum contributions $\propto 1$ to
obtain
\bea\label{eq:pol_ij2}
\Pi^{(1)}_{ij}(\bm q,\omega_n)
=-2e^{*2}\delta_{ij}\int \frac{d\bm p}{(2\pi)^3}
\biggl[\frac{f^+(E_p)+f^-(E_{p})}{E_p}\bigg].
\eea
In the remaining part of the polarization tensor we set
$(p_ip'_j+p_jp'_i)/(E_pE_{p-q}) \simeq {2p_ip_j}/{E_p^2}$ to find
\bea\label{eq:pol_ij4}
\Pi^{(2)}_{ij}(\bm q,\omega_n) 
&=&-2e^{*2}\int \frac{d\bm p}{(2\pi)^3}\frac{p_ip_j}{E_p^2}
\bigg\{
\bigg[
\frac{f^+(E_p)-f^+(E_{p-q})}
{E_p-E_{p-q}-i\omega_n}+
\frac{f^+(-E_p)-f^+(-E_{p-q})}
{-E_p+E_{p-q}-i\omega_n}\bigg]\nonumber\\
&-&
\biggl[
\frac{f^+(E_p)-f^+(-E_{p-q})}
{E_p+E_{p-q}-i\omega_n}+
\frac{f^+(-E_p)-f^+(E_{p-q})}
{-E_p-E_{p-q}-i\omega_n}\bigg]\bigg\}. 
\eea  
We further approximate 
\bea
&&E_{p-q}=E_p - \frac{(\vecp\cdot
  \vecq)}{E_p},\\
&&f^+(E_p)-f^+(E_{p-q})=\frac{(\vecp\cdot
  \vecq)}{E_p} \frac{\partial f^+(E_p)}{\partial E_p},\\
&&f^+(-E_p)-f^+(-E_{p-q}) = - \frac{(\vecp\cdot 
  \vecq)}{E_p} \frac{\partial f^-(E_p)}{\partial E_p},
\eea
and drop the vacuum terms to obtain
\bea\label{eq:pol_ij5}
\Pi^{(2)}_{ij}(\bm q,\omega_n) 
&=&-2e^{*2}\int \frac{d\bm p}{(2\pi)^3}\frac{p_ip_j}{E_p^2}
\bigg\{
\left[\frac{\partial f^+(E_p)}{\partial E_p}+
\frac{\partial f^-(E_p)}{\partial E_p}\right]
+\nonumber\\&+&
\frac{i\omega E_p}{(\vecp\cdot \vecq)-i\omega_nE_p}
\left[\frac{\partial f^+(E_p)}{\partial E_p}+
\frac{\partial f^-(E_p)}{\partial E_p}\right]-\frac{f^+(E_p)+f^-(E_{p})}{E_p}\bigg\}.
\eea
Now we add Eq.~\eqref{eq:pol_ij2} to this to obtain 
\bea\label{eq:pol_ij6}
\Pi_{ij}(\bm q,\omega_n) 
&=&-2e^{*2}\int\frac{p^2dp}{2\pi^2}\int \frac{d\Omega}{4\pi}\bigg\{
\delta_{ij}
\biggl[\frac{f^+(E_p)+f^-(E_{p})}{E_p}\bigg]
-\frac{p_ip_j}{E_p^2}\left[\frac{f^+(E_p)+f^-(E_{p})}{E_p}\right]\nonumber\\
&+&\frac{p_ip_j}{E_p^2}
\left[\frac{\partial f^+(E_p)}{\partial E_p}+
\frac{\partial f^-(E_p)}{\partial E_p}\right]
-\frac{p_ip_j}{E_p^2}\frac{i\omega E_p}{  i\omega_nE_p-(\vecp\cdot \vecq)}
\left[\frac{\partial f^+(E_p)}{\partial E_p}+
\frac{\partial f^-(E_p)}{\partial E_p}\right]\bigg\}. 
\eea
In the first three terms in Eq.~\eqref{eq:pol_ij6} the angular and
radial integrals separate. For the angular part we have 
\bea
\int \frac{d\Omega}{4\pi}\frac{p_ip_j}{p^2} = \frac{\delta_{ij}}{3}.
\eea
By partial integration we find 
\bea 
\label{eq:ex1_1}
\int d\ep \frac{p^3}{\ep}\frac{d}{d\ep}[f^+(\ep)+f^-(\ep)] 
=- 3\int [f^+(\ep)+f^-(\ep)]
\bigg(1-\frac{p^2}{3\ep^2} \bigg)pd \ep
\eea 
which implies that the sum of these terms vanishes.  For the remainder
we find 
\bea 
\Pi_{ij}(\bm q,\omega)&=&
\frac{4e^2}{\pi}\int p^2dp 
\left[\frac{\partial f^+(E_p)}{\partial E_p}+
\frac{\partial f^-(E_p)}{\partial E_p}\right]v^2
\int \frac{d\Omega}{4\pi}
n_in_j\frac{\omega }{  \omega + i\delta -v(\vecn\cdot \vecq)},
\eea
where $v=(\partial E_p/\partial p )= p/E_p$ and in the last term we
have performed the analytical continuation
$i\omega_n \to \omega + i\delta$.  The spatial part of the
polarization tensor can be decomposed as
\bea \label{eq:decomp_0}
\Pi_{ij}(\bm q,\omega)  =P_{ij}^T \Pi_T(\bm q,\omega)+P_{ij}^L
\Pi_L(\bm q,\omega).
\eea
Contracting Eq. \eqref{eq:decomp_0} with $\delta_{ij}$ and using 
Eqs.~\eqref{eq:trans} and \eqref{eq:long} we find 
\bea\label{eq:decomp1}
2 \Pi_T(\bm q,\omega)+\frac{\omega^2}{\omega^2-q^2}
 \Pi_L(\bm q,\omega) =
\left(-\frac{4e^2}{\pi}\right)\int p^2dp 
\left[\frac{\partial f^+(E_p)}{\partial E_p}+
\frac{\partial f^-(E_p)}{\partial E_p}\right]
v^2L_v(\bm q, \omega),
\eea
where
\bea\label{eq:L_v}
L_v(\bm q, \omega)\equiv\int \frac{d\Omega}{4\pi}
\frac{ \omega}{\omega+i\delta-v(\vecn\cdot \vecq)}=\frac{x}{2v}
\log\frac{x+v}{x-v},\qquad x\equiv\omega/q.
\eea
Next contract Eq. \eqref{eq:decomp_0} with $q_iq_j$ 
to find (note that $q_iq_jP_{ij}^T = 0$)
\bea\label{eq:decomp2}
\frac{q^2}{\omega^2-q^2}\Pi^L(\bm  q,\omega) = 
\left(-\frac{4e^2}{\pi}\right)\int p^2dp 
\left[\frac{\partial f^+(E_p)}{\partial E_p}+
\frac{\partial f^-(E_p)}{\partial E_p}\right]
[L_v(\bm  q,\omega)-1].
\eea
Using Eqs.~\eqref{eq:decomp1}-\eqref{eq:decomp2} 
 we obtain Eqs.~\eqref{eq:pol_final_l} and \eqref{eq:pol_final_t} of the main text.
\end{widetext}

\bibliographystyle{JHEP}

\providecommand{\href}[2]{#2}\begingroup\raggedright\endgroup

\newpage
\setlength{\textheight}{25cm}
\begin{widetext}
\section*{Supplemental material}

Below we present the numerical tables for the conductivities log$_{10}\sigma$, log$_{10}\sigma_0$ and log$_{10}\sigma_1$ (in units of s$^{-1}$) for various values of magnetic field (in units of $10^{12}$ G) for a set of values of density [g cm$^{-3}$] and temperature [MeV].  The tables are provided for three types of composition of matter: $\isotope[12]{C}$ nuclei, $\isotope[56]{Fe}$ nuclei, and density-dependent composition as indicated in Fig.~2.

\begin{table}
\begin{center}
\caption {$\log\sigma$ for $\isotope[12]{C}$} \label{tab:sigma_C}
{\tiny
\begin{tabular}{ cccccccccccc }
\hline
  $\log\rho$  &  $\log T$=-1.0  &  $\log T$=-0.8  &  $\log T$=-0.6  &  $\log T$=-0.4  &  $\log T$=-0.2  &  $\log T$=0.0  &  $\log T$=0.2  &  $\log T$=0.4  &  $\log T$=0.6  &  $\log T$=0.8  &  $\log T$=1.0  \\
  \hline
6.0  &  20.745  &  20.853  &  20.973  &  21.100  &  21.222  &  21.356  &  21.498  &  21.646  &  21.798  &  21.957  &  22.123   \\
 6.2  &  20.816  &  20.905  &  21.013  &  21.131  &  21.251  &  21.378  &  21.518  &  21.663  &  21.814  &  21.970  &  22.136   \\
 6.4  &  20.898  &  20.965  &  21.057  &  21.165  &  21.281  &  21.401  &  21.538  &  21.681  &  21.830  &  21.985  &  22.149   \\
 6.6  &  20.990  &  21.033  &  21.106  &  21.202  &  21.313  &  21.426  &  21.559  &  21.699  &  21.846  &  22.000  &  22.162   \\
 6.8  &  21.089  &  21.110  &  21.163  &  21.244  &  21.346  &  21.452  &  21.581  &  21.719  &  21.864  &  22.015  &  22.176   \\
 7.0  &  21.192  &  21.193  &  21.226  &  21.290  &  21.381  &  21.482  &  21.605  &  21.739  &  21.882  &  22.031  &  22.190   \\
 7.2  &  21.296  &  21.283  &  21.296  &  21.342  &  21.420  &  21.518  &  21.630  &  21.760  &  21.900  &  22.048  &  22.205   \\
\hline
\end{tabular}
} 
\caption {$\log\sigma_0$ for $\isotope[12]{C}$ at $B_{12}=1$} \label{tab:sigma0_b12_C}
{\tiny
\begin{tabular}{ cccccccccccc }
\hline
  $\log\rho$  &  $\log T$=-1.0  &  $\log T$=-0.8  &  $\log T$=-0.6  &  $\log T$=-0.4  &  $\log T$=-0.2  &  $\log T$=0.0  &  $\log T$=0.2  &  $\log T$=0.4  &  $\log T$=0.6  &  $\log T$=0.8  &  $\log T$=1.0  \\
  \hline
6.0  &  16.648  &  16.551  &  16.423  &  16.279  &  16.132  &  15.987  &  15.843  &  15.696  &  15.546  &  15.390  &  15.227   \\
 6.2  &  16.947  &  16.878  &  16.771  &  16.641  &  16.503  &  16.363  &  16.222  &  16.078  &  15.929  &  15.776  &  15.613   \\
 6.4  &  17.238  &  17.195  &  17.112  &  16.999  &  16.871  &  16.736  &  16.599  &  16.458  &  16.312  &  16.160  &  15.999   \\
 6.6  &  17.522  &  17.504  &  17.445  &  17.352  &  17.236  &  17.108  &  16.976  &  16.838  &  16.694  &  16.544  &  16.385   \\
 6.8  &  17.805  &  17.806  &  17.771  &  17.699  &  17.597  &  17.478  &  17.351  &  17.216  &  17.075  &  16.928  &  16.770   \\
 7.0  &  18.090  &  18.104  &  18.091  &  18.040  &  17.955  &  17.845  &  17.724  &  17.594  &  17.455  &  17.310  &  17.155   \\
 7.2  &  18.378  &  18.402  &  18.405  &  18.375  &  18.307  &  18.211  &  18.096  &  17.970  &  17.835  &  17.692  &  17.539   \\
\hline
\end{tabular}
} 
\end{center}
\end{table}
\begin{table}
\begin{center}
\caption {$\log\sigma_0$ for $\isotope[12]{C}$ at $B_{12}=10$} \label{tab:sigma0_b13_C}
{\tiny
\begin{tabular}{ cccccccccccc }
\hline
  $\log\rho$  &  $\log T$=-1.0  &  $\log T$=-0.8  &  $\log T$=-0.6  &  $\log T$=-0.4  &  $\log T$=-0.2  &  $\log T$=0.0  &  $\log T$=0.2  &  $\log T$=0.4  &  $\log T$=0.6  &  $\log T$=0.8  &  $\log T$=1.0  \\
  \hline
6.0  &  14.648  &  14.551  &  14.423  &  14.280  &  14.133  &  13.988  &  13.842  &  13.696  &  13.546  &  13.390  &  13.227   \\
 6.2  &  14.948  &  14.878  &  14.771  &  14.642  &  14.503  &  14.363  &  14.221  &  14.078  &  13.929  &  13.776  &  13.613   \\
 6.4  &  15.238  &  15.195  &  15.112  &  14.999  &  14.871  &  14.736  &  14.599  &  14.458  &  14.312  &  14.160  &  13.999   \\
 6.6  &  15.523  &  15.504  &  15.446  &  15.352  &  15.236  &  15.108  &  14.975  &  14.838  &  14.694  &  14.544  &  14.385   \\
 6.8  &  15.806  &  15.806  &  15.772  &  15.700  &  15.598  &  15.478  &  15.351  &  15.216  &  15.075  &  14.928  &  14.770   \\
 7.0  &  16.090  &  16.105  &  16.092  &  16.041  &  15.955  &  15.846  &  15.724  &  15.594  &  15.455  &  15.310  &  15.155   \\
 7.2  &  16.379  &  16.403  &  16.406  &  16.376  &  16.308  &  16.212  &  16.096  &  15.970  &  15.835  &  15.692  &  15.539   \\
\hline
\end{tabular}
} 
\caption {$\log\sigma_0$ for $\isotope[12]{C}$ at $B_{12}=100$} \label{tab:sigma0_b14_C}
{\tiny
\begin{tabular}{ cccccccccccc }
\hline
  $\log\rho$  &  $\log T$=-1.0  &  $\log T$=-0.8  &  $\log T$=-0.6  &  $\log T$=-0.4  &  $\log T$=-0.2  &  $\log T$=0.0  &  $\log T$=0.2  &  $\log T$=0.4  &  $\log T$=0.6  &  $\log T$=0.8  &  $\log T$=1.0  \\
  \hline
6.0  &  12.648  &  12.551  &  12.423  &  12.280  &  12.133  &  11.988  &  11.843  &  11.696  &  11.546  &  11.390  &  11.227   \\
 6.2  &  12.948  &  12.878  &  12.771  &  12.642  &  12.503  &  12.363  &  12.222  &  12.078  &  11.929  &  11.776  &  11.613   \\
 6.4  &  13.238  &  13.195  &  13.112  &  12.999  &  12.871  &  12.736  &  12.599  &  12.458  &  12.312  &  12.160  &  11.999   \\
 6.6  &  13.523  &  13.504  &  13.446  &  13.352  &  13.236  &  13.108  &  12.976  &  12.838  &  12.694  &  12.544  &  12.385   \\
 6.8  &  13.806  &  13.806  &  13.772  &  13.700  &  13.598  &  13.478  &  13.351  &  13.216  &  13.075  &  12.928  &  12.770   \\
 7.0  &  14.090  &  14.105  &  14.092  &  14.041  &  13.955  &  13.846  &  13.724  &  13.594  &  13.455  &  13.310  &  13.155   \\
 7.2  &  14.379  &  14.403  &  14.406  &  14.376  &  14.308  &  14.212  &  14.096  &  13.970  &  13.835  &  13.692  &  13.539   \\
\hline
\end{tabular}
} 
\end{center}
\end{table}
\begin{table}
\begin{center}
\caption {$\log\sigma_1$ for $\isotope[12]{C}$ at $B_{12}=1$} \label{tab:sigma1_b12_C}
{\tiny
\begin{tabular}{ cccccccccccc }
\hline
  $\log\rho$  &  $\log T$=-1.0  &  $\log T$=-0.8  &  $\log T$=-0.6  &  $\log T$=-0.4  &  $\log T$=-0.2  &  $\log T$=0.0  &  $\log T$=0.2  &  $\log T$=0.4  &  $\log T$=0.6  &  $\log T$=0.8  &  $\log T$=1.0  \\
  \hline
6.0  &  18.633  &  18.633  &  18.633  &  18.633  &  18.629  &  18.629  &  18.629  &  18.629  &  18.629  &  18.629  &  18.629   \\
 6.2  &  18.833  &  18.833  &  18.833  &  18.833  &  18.831  &  18.829  &  18.829  &  18.829  &  18.829  &  18.829  &  18.829   \\
 6.4  &  19.033  &  19.033  &  19.033  &  19.033  &  19.032  &  19.029  &  19.029  &  19.029  &  19.029  &  19.029  &  19.029   \\
 6.6  &  19.233  &  19.233  &  19.233  &  19.233  &  19.233  &  19.229  &  19.229  &  19.229  &  19.229  &  19.229  &  19.229   \\
 6.8  &  19.433  &  19.433  &  19.433  &  19.433  &  19.433  &  19.429  &  19.429  &  19.429  &  19.429  &  19.429  &  19.429   \\
 7.0  &  19.633  &  19.633  &  19.633  &  19.633  &  19.633  &  19.629  &  19.629  &  19.629  &  19.629  &  19.629  &  19.629   \\
 7.2  &  19.833  &  19.833  &  19.833  &  19.833  &  19.833  &  19.831  &  19.829  &  19.829  &  19.829  &  19.829  &  19.829   \\
\hline
\end{tabular}
} 
\caption {$\log\sigma_1$ for $\isotope[12]{C}$ at $B_{12}=10$} \label{tab:sigma1_b13_C}
{\tiny
\begin{tabular}{ cccccccccccc }
\hline
  $\log\rho$  &  $\log T$=-1.0  &  $\log T$=-0.8  &  $\log T$=-0.6  &  $\log T$=-0.4  &  $\log T$=-0.2  &  $\log T$=0.0  &  $\log T$=0.2  &  $\log T$=0.4  &  $\log T$=0.6  &  $\log T$=0.8  &  $\log T$=1.0  \\
  \hline
6.0  &  17.633  &  17.633  &  17.633  &  17.633  &  17.629  &  17.629  &  17.629  &  17.629  &  17.629  &  17.629  &  17.629   \\
 6.2  &  17.833  &  17.833  &  17.833  &  17.833  &  17.831  &  17.829  &  17.829  &  17.829  &  17.829  &  17.829  &  17.829   \\
 6.4  &  18.033  &  18.033  &  18.033  &  18.033  &  18.032  &  18.029  &  18.029  &  18.029  &  18.029  &  18.029  &  18.029   \\
 6.6  &  18.233  &  18.233  &  18.233  &  18.233  &  18.233  &  18.229  &  18.229  &  18.229  &  18.229  &  18.229  &  18.229   \\
 6.8  &  18.433  &  18.433  &  18.433  &  18.433  &  18.433  &  18.429  &  18.429  &  18.429  &  18.429  &  18.429  &  18.429   \\
 7.0  &  18.633  &  18.633  &  18.633  &  18.633  &  18.633  &  18.629  &  18.629  &  18.629  &  18.629  &  18.629  &  18.629   \\
 7.2  &  18.833  &  18.833  &  18.833  &  18.833  &  18.833  &  18.832  &  18.829  &  18.829  &  18.829  &  18.829  &  18.829   \\
\hline
\end{tabular}
} 
\end{center}
\end{table}
\begin{table}
\begin{center}
\caption {$\log\sigma_1$ for $\isotope[12]{C}$ at $B_{12}=100$} \label{tab:sigma1_b14_C}
{\tiny
\begin{tabular}{ cccccccccccc }
\hline
  $\log\rho$  &  $\log T$=-1.0  &  $\log T$=-0.8  &  $\log T$=-0.6  &  $\log T$=-0.4  &  $\log T$=-0.2  &  $\log T$=0.0  &  $\log T$=0.2  &  $\log T$=0.4  &  $\log T$=0.6  &  $\log T$=0.8  &  $\log T$=1.0  \\
  \hline
6.0  &  16.633  &  16.633  &  16.633  &  16.633  &  16.629  &  16.629  &  16.629  &  16.629  &  16.629  &  16.629  &  16.629   \\
 6.2  &  16.833  &  16.833  &  16.833  &  16.833  &  16.831  &  16.829  &  16.829  &  16.829  &  16.829  &  16.829  &  16.829   \\
 6.4  &  17.033  &  17.033  &  17.033  &  17.033  &  17.032  &  17.029  &  17.029  &  17.029  &  17.029  &  17.029  &  17.029   \\
 6.6  &  17.233  &  17.233  &  17.233  &  17.233  &  17.233  &  17.229  &  17.229  &  17.229  &  17.229  &  17.229  &  17.229   \\
 6.8  &  17.433  &  17.433  &  17.433  &  17.433  &  17.433  &  17.429  &  17.429  &  17.429  &  17.429  &  17.429  &  17.429   \\
 7.0  &  17.633  &  17.633  &  17.633  &  17.633  &  17.633  &  17.629  &  17.629  &  17.629  &  17.629  &  17.629  &  17.629   \\
 7.2  &  17.833  &  17.833  &  17.833  &  17.833  &  17.833  &  17.832  &  17.829  &  17.829  &  17.829  &  17.829  &  17.829   \\
\hline
\end{tabular}
} 
\caption {$\log\sigma$ for $\isotope[56]{Fe}$} \label{tab:sigma_Fe}
{\tiny
\begin{tabular}{ cccccccccccc }
\hline
  $\log\rho$  &  $\log T$=-1.0  &  $\log T$=-0.8  &  $\log T$=-0.6  &  $\log T$=-0.4  &  $\log T$=-0.2  &  $\log T$=0.0  &  $\log T$=0.2  &  $\log T$=0.4  &  $\log T$=0.6  &  $\log T$=0.8  &  $\log T$=1.0  \\
  \hline
6.0  &  20.144  &  20.260  &  20.375  &  20.506  &  20.628  &  20.762  &  20.902  &  21.048  &  21.200  &  21.357  &  21.522   \\
 6.2  &  20.213  &  20.310  &  20.413  &  20.537  &  20.658  &  20.785  &  20.923  &  21.066  &  21.216  &  21.372  &  21.536   \\
 6.4  &  20.293  &  20.368  &  20.456  &  20.570  &  20.689  &  20.809  &  20.944  &  21.085  &  21.233  &  21.388  &  21.550   \\
 6.6  &  20.382  &  20.435  &  20.517  &  20.607  &  20.720  &  20.834  &  20.966  &  21.104  &  21.250  &  21.403  &  21.565   \\
 6.8  &  20.477  &  20.510  &  20.572  &  20.647  &  20.753  &  20.861  &  20.989  &  21.125  &  21.268  &  21.420  &  21.580   \\
 7.0  &  20.576  &  20.592  &  20.634  &  20.692  &  20.788  &  20.889  &  21.013  &  21.146  &  21.287  &  21.437  &  21.595   \\
 7.2  &  20.675  &  20.680  &  20.702  &  20.756  &  20.826  &  20.925  &  21.038  &  21.168  &  21.307  &  21.454  &  21.611   \\
\hline
\end{tabular}
} 
\end{center}
\end{table}
\begin{table}
\begin{center}
\caption {$\log\sigma_0$ for $\isotope[56]{Fe}$ at $B_{12}=1$} \label{tab:sigma0_b12_Fe}
{\tiny
\begin{tabular}{ cccccccccccc }
\hline
  $\log\rho$  &  $\log T$=-1.0  &  $\log T$=-0.8  &  $\log T$=-0.6  &  $\log T$=-0.4  &  $\log T$=-0.2  &  $\log T$=0.0  &  $\log T$=0.2  &  $\log T$=0.4  &  $\log T$=0.6  &  $\log T$=0.8  &  $\log T$=1.0  \\
  \hline
6.0  &  17.169  &  17.065  &  16.945  &  16.800  &  16.654  &  16.512  &  16.370  &  16.226  &  16.078  &  15.925  &  15.766   \\
 6.2  &  17.472  &  17.395  &  17.295  &  17.161  &  17.023  &  16.886  &  16.748  &  16.607  &  16.460  &  16.309  &  16.151   \\
 6.4  &  17.768  &  17.715  &  17.637  &  17.519  &  17.391  &  17.258  &  17.124  &  16.986  &  16.842  &  16.692  &  16.536   \\
 6.6  &  18.057  &  18.026  &  17.957  &  17.872  &  17.755  &  17.629  &  17.499  &  17.364  &  17.223  &  17.075  &  16.920   \\
 6.8  &  18.345  &  18.331  &  18.285  &  18.219  &  18.115  &  17.997  &  17.873  &  17.742  &  17.603  &  17.457  &  17.304   \\
 7.0  &  18.634  &  18.631  &  18.605  &  18.560  &  18.471  &  18.362  &  18.245  &  18.117  &  17.982  &  17.839  &  17.687   \\
 7.2  &  18.926  &  18.930  &  18.919  &  18.879  &  18.821  &  18.725  &  18.614  &  18.492  &  18.360  &  18.219  &  18.070   \\
\hline
\end{tabular}
} 
\end{center}
\begin{center}
\caption {$\log\sigma_0$ for $\isotope[56]{Fe}$ at $B_{12}=10$} \label{tab:sigma0_b13_Fe}
{\tiny
\begin{tabular}{ cccccccccccc }
\hline
  $\log\rho$  &  $\log T$=-1.0  &  $\log T$=-0.8  &  $\log T$=-0.6  &  $\log T$=-0.4  &  $\log T$=-0.2  &  $\log T$=0.0  &  $\log T$=0.2  &  $\log T$=0.4  &  $\log T$=0.6  &  $\log T$=0.8  &  $\log T$=1.0  \\
  \hline
6.0  &  15.171  &  15.068  &  14.948  &  14.801  &  14.655  &  14.512  &  14.370  &  14.226  &  14.078  &  13.925  &  13.766   \\
 6.2  &  15.475  &  15.398  &  15.298  &  15.163  &  15.024  &  14.886  &  14.748  &  14.607  &  14.460  &  14.309  &  14.151   \\
 6.4  &  15.771  &  15.718  &  15.641  &  15.522  &  15.392  &  15.259  &  15.125  &  14.986  &  14.842  &  14.692  &  14.536   \\
 6.6  &  16.061  &  16.031  &  15.961  &  15.875  &  15.757  &  15.630  &  15.500  &  15.365  &  15.223  &  15.075  &  14.920   \\
 6.8  &  16.350  &  16.336  &  16.290  &  16.224  &  16.118  &  15.999  &  15.874  &  15.742  &  15.603  &  15.457  &  15.304   \\
 7.0  &  16.640  &  16.638  &  16.612  &  16.568  &  16.476  &  16.366  &  16.246  &  16.119  &  15.983  &  15.839  &  15.688   \\
 7.2  &  16.934  &  16.939  &  16.929  &  16.889  &  16.830  &  16.731  &  16.617  &  16.494  &  16.361  &  16.220  &  16.070   \\
\hline
\end{tabular}
} 
\end{center}
\end{table}
\begin{table}
\begin{center}
\caption {$\log\sigma_0$ for $\isotope[56]{Fe}$ at $B_{12}=100$} \label{tab:sigma0_b14_Fe}
{\tiny
\begin{tabular}{ cccccccccccc }
\hline
  $\log\rho$  &  $\log T$=-1.0  &  $\log T$=-0.8  &  $\log T$=-0.6  &  $\log T$=-0.4  &  $\log T$=-0.2  &  $\log T$=0.0  &  $\log T$=0.2  &  $\log T$=0.4  &  $\log T$=0.6  &  $\log T$=0.8  &  $\log T$=1.0  \\
  \hline
6.0  &  13.171  &  13.068  &  12.948  &  12.801  &  12.655  &  12.512  &  12.370  &  12.226  &  12.078  &  11.925  &  11.766   \\
 6.2  &  13.475  &  13.398  &  13.298  &  13.163  &  13.025  &  12.886  &  12.748  &  12.607  &  12.460  &  12.309  &  12.151   \\
 6.4  &  13.771  &  13.718  &  13.641  &  13.522  &  13.392  &  13.259  &  13.125  &  12.986  &  12.842  &  12.692  &  12.536   \\
 6.6  &  14.061  &  14.031  &  13.961  &  13.876  &  13.757  &  13.630  &  13.500  &  13.365  &  13.223  &  13.075  &  12.920   \\
 6.8  &  14.350  &  14.336  &  14.290  &  14.224  &  14.118  &  13.999  &  13.874  &  13.742  &  13.603  &  13.457  &  13.304   \\
 7.0  &  14.640  &  14.638  &  14.612  &  14.568  &  14.476  &  14.366  &  14.246  &  14.119  &  13.983  &  13.839  &  13.688   \\
 7.2  &  14.934  &  14.939  &  14.929  &  14.889  &  14.830  &  14.731  &  14.617  &  14.494  &  14.361  &  14.220  &  14.070   \\
\hline
\end{tabular}
} 
\caption {$\log\sigma_1$ for $\isotope[56]{Fe}$ at $B_{12}=1$} \label{tab:sigma0_b12_Fe}
{\tiny
\begin{tabular}{ cccccccccccc }
\hline
  $\log\rho$  &  $\log T$=-1.0  &  $\log T$=-0.8  &  $\log T$=-0.6  &  $\log T$=-0.4  &  $\log T$=-0.2  &  $\log T$=0.0  &  $\log T$=0.2  &  $\log T$=0.4  &  $\log T$=0.6  &  $\log T$=0.8  &  $\log T$=1.0  \\
  \hline
6.0  &  18.600  &  18.601  &  18.601  &  18.601  &  18.597  &  18.596  &  18.597  &  18.597  &  18.597  &  18.597  &  18.597   \\
 6.2  &  18.800  &  18.800  &  18.801  &  18.801  &  18.798  &  18.796  &  18.797  &  18.797  &  18.797  &  18.797  &  18.797   \\
 6.4  &  18.999  &  19.000  &  19.000  &  19.000  &  18.999  &  18.996  &  18.996  &  18.997  &  18.997  &  18.997  &  18.997   \\
 6.6  &  19.199  &  19.199  &  19.199  &  19.200  &  19.200  &  19.196  &  19.196  &  19.197  &  19.197  &  19.197  &  19.197   \\
 6.8  &  19.398  &  19.398  &  19.398  &  19.399  &  19.399  &  19.396  &  19.396  &  19.396  &  19.397  &  19.397  &  19.397   \\
 7.0  &  19.596  &  19.596  &  19.596  &  19.597  &  19.598  &  19.595  &  19.596  &  19.596  &  19.596  &  19.597  &  19.597   \\
 7.2  &  19.793  &  19.793  &  19.793  &  19.794  &  19.795  &  19.796  &  19.794  &  19.795  &  19.796  &  19.796  &  19.796   \\
\hline
\end{tabular}
} 
\end{center}
\end{table}
\begin{table}
\begin{center}
\caption {$\log\sigma_1$ for $\isotope[56]{Fe}$ at $B_{12}=10$} \label{tab:sigma0_b13_Fe}
{\tiny
\begin{tabular}{ cccccccccccc }
\hline
  $\log\rho$  &  $\log T$=-1.0  &  $\log T$=-0.8  &  $\log T$=-0.6  &  $\log T$=-0.4  &  $\log T$=-0.2  &  $\log T$=0.0  &  $\log T$=0.2  &  $\log T$=0.4  &  $\log T$=0.6  &  $\log T$=0.8  &  $\log T$=1.0  \\
  \hline
6.0  &  17.601  &  17.601  &  17.601  &  17.601  &  17.597  &  17.596  &  17.597  &  17.597  &  17.597  &  17.597  &  17.597   \\
 6.2  &  17.801  &  17.801  &  17.801  &  17.801  &  17.799  &  17.796  &  17.797  &  17.797  &  17.797  &  17.797  &  17.797   \\
 6.4  &  18.001  &  18.001  &  18.001  &  18.001  &  18.000  &  17.996  &  17.997  &  17.997  &  17.997  &  17.997  &  17.997   \\
 6.6  &  18.201  &  18.201  &  18.201  &  18.201  &  18.201  &  18.196  &  18.197  &  18.197  &  18.197  &  18.197  &  18.197   \\
 6.8  &  18.401  &  18.401  &  18.401  &  18.401  &  18.401  &  18.396  &  18.397  &  18.397  &  18.397  &  18.397  &  18.397   \\
 7.0  &  18.601  &  18.601  &  18.601  &  18.601  &  18.601  &  18.596  &  18.597  &  18.597  &  18.597  &  18.597  &  18.597   \\
 7.2  &  18.801  &  18.801  &  18.801  &  18.801  &  18.801  &  18.799  &  18.797  &  18.797  &  18.797  &  18.797  &  18.797   \\
\hline
\end{tabular}
} 
\caption {$\log\sigma_1$ for $\isotope[56]{Fe}$ at $B_{12}=100$} \label{tab:sigma0_b14_Fe}
{\tiny
\begin{tabular}{ cccccccccccc }
\hline
  $\log\rho$  &  $\log T$=-1.0  &  $\log T$=-0.8  &  $\log T$=-0.6  &  $\log T$=-0.4  &  $\log T$=-0.2  &  $\log T$=0.0  &  $\log T$=0.2  &  $\log T$=0.4  &  $\log T$=0.6  &  $\log T$=0.8  &  $\log T$=1.0  \\
  \hline
6.0  &  16.601  &  16.601  &  16.601  &  16.601  &  16.597  &  16.596  &  16.597  &  16.597  &  16.597  &  16.597  &  16.597   \\
 6.2  &  16.801  &  16.801  &  16.801  &  16.801  &  16.799  &  16.796  &  16.797  &  16.797  &  16.797  &  16.797  &  16.797   \\
 6.4  &  17.001  &  17.001  &  17.001  &  17.001  &  17.000  &  16.996  &  16.997  &  16.997  &  16.997  &  16.997  &  16.997   \\
 6.6  &  17.201  &  17.201  &  17.201  &  17.201  &  17.201  &  17.196  &  17.197  &  17.197  &  17.197  &  17.197  &  17.197   \\
 6.8  &  17.401  &  17.401  &  17.401  &  17.401  &  17.401  &  17.396  &  17.397  &  17.397  &  17.397  &  17.397  &  17.397   \\
 7.0  &  17.601  &  17.601  &  17.601  &  17.601  &  17.601  &  17.596  &  17.597  &  17.597  &  17.597  &  17.597  &  17.597   \\
 7.2  &  17.801  &  17.801  &  17.801  &  17.801  &  17.801  &  17.799  &  17.797  &  17.797  &  17.797  &  17.797  &  17.797   \\
\hline
\end{tabular}
} 
\end{center}
\end{table}
\begin{table}
\begin{center}
\caption {$\log\sigma$ for density-dependent composition} \label{tab:sigma_comp}
{\tiny
\begin{tabular}{ cccccccccccc }
\hline
  $\log\rho$  &  $\log T$=-1.0  &  $\log T$=-0.8  &  $\log T$=-0.6  &  $\log T$=-0.4  &  $\log T$=-0.2  &  $\log T$=0.0  &  $\log T$=0.2  &  $\log T$=0.4  &  $\log T$=0.6  &  $\log T$=0.8  &  $\log T$=1.0  \\
  \hline
6.0  &  20.144  &  20.260  &  20.376  &  20.506  &  20.628  &  20.762  &  20.902  &  21.048  &  21.200  &  21.357  &  21.522   \\
 6.2  &  20.213  &  20.310  &  20.413  &  20.537  &  20.658  &  20.785  &  20.923  &  21.066  &  21.216  &  21.372  &  21.536   \\
 6.4  &  20.293  &  20.368  &  20.456  &  20.570  &  20.689  &  20.809  &  20.944  &  21.085  &  21.233  &  21.388  &  21.550   \\
 6.6  &  20.382  &  20.435  &  20.517  &  20.607  &  20.720  &  20.834  &  20.966  &  21.104  &  21.250  &  21.403  &  21.565   \\
 6.8  &  20.477  &  20.510  &  20.572  &  20.647  &  20.753  &  20.861  &  20.989  &  21.125  &  21.268  &  21.420  &  21.580   \\
 7.0  &  20.536  &  20.555  &  20.598  &  20.671  &  20.755  &  20.857  &  20.981  &  21.114  &  21.256  &  21.406  &  21.565   \\
 7.2  &  20.633  &  20.641  &  20.666  &  20.721  &  20.792  &  20.893  &  21.006  &  21.136  &  21.276  &  21.424  &  21.581   \\
\hline
\end{tabular}
} 
\end{center}
\begin{center}
\caption {$\log\sigma_0$ for density-dependent composition at $B_{12}=1$} \label{tab:sigma0_b12_comp}
{\tiny
\begin{tabular}{ cccccccccccc }
\hline
  $\log\rho$  &  $\log T$=-1.0  &  $\log T$=-0.8  &  $\log T$=-0.6  &  $\log T$=-0.4  &  $\log T$=-0.2  &  $\log T$=0.0  &  $\log T$=0.2  &  $\log T$=0.4  &  $\log T$=0.6  &  $\log T$=0.8  &  $\log T$=1.0  \\
  \hline
6.0  &  17.169  &  17.065  &  16.945  &  16.800  &  16.654  &  16.512  &  16.370  &  16.226  &  16.078  &  15.925  &  15.765   \\
 6.2  &  17.472  &  17.395  &  17.294  &  17.161  &  17.023  &  16.885  &  16.748  &  16.607  &  16.460  &  16.309  &  16.151   \\
 6.4  &  17.768  &  17.715  &  17.637  &  17.519  &  17.390  &  17.258  &  17.124  &  16.986  &  16.842  &  16.692  &  16.536   \\
 6.6  &  18.057  &  18.026  &  17.957  &  17.872  &  17.754  &  17.628  &  17.499  &  17.364  &  17.223  &  17.075  &  16.920   \\
 6.8  &  18.345  &  18.331  &  18.285  &  18.219  &  18.115  &  17.997  &  17.873  &  17.741  &  17.603  &  17.457  &  17.304   \\
 7.0  &  18.650  &  18.645  &  18.617  &  18.555  &  18.480  &  18.370  &  18.252  &  18.124  &  17.989  &  17.845  &  17.694   \\
 7.2  &  18.943  &  18.943  &  18.931  &  18.889  &  18.830  &  18.733  &  18.621  &  18.499  &  18.367  &  18.226  &  18.076   \\
\hline
\end{tabular}
} 
\end{center}
\end{table}
\begin{table}
\begin{center}
\caption {$\log\sigma_0$ for density-dependent composition at $B_{12}=10$} \label{tab:sigma0_b13_comp}
{\tiny
\begin{tabular}{ cccccccccccc }
\hline
  $\log\rho$  &  $\log T$=-1.0  &  $\log T$=-0.8  &  $\log T$=-0.6  &  $\log T$=-0.4  &  $\log T$=-0.2  &  $\log T$=0.0  &  $\log T$=0.2  &  $\log T$=0.4  &  $\log T$=0.6  &  $\log T$=0.8  &  $\log T$=1.0  \\
  \hline
6.0  &  15.171  &  15.068  &  14.948  &  14.801  &  14.654  &  14.512  &  14.370  &  14.226  &  14.078  &  13.925  &  13.765   \\
 6.2  &  15.475  &  15.398  &  15.297  &  15.163  &  15.024  &  14.886  &  14.748  &  14.607  &  14.461  &  14.309  &  14.151   \\
 6.4  &  15.771  &  15.718  &  15.641  &  15.521  &  15.392  &  15.259  &  15.124  &  14.986  &  14.842  &  14.692  &  14.536   \\
 6.6  &  16.061  &  16.031  &  15.961  &  15.875  &  15.757  &  15.630  &  15.500  &  15.364  &  15.223  &  15.075  &  14.920   \\
 6.8  &  16.350  &  16.336  &  16.290  &  16.224  &  16.118  &  15.999  &  15.874  &  15.742  &  15.603  &  15.457  &  15.304   \\
 7.0  &  16.657  &  16.652  &  16.625  &  16.563  &  16.486  &  16.374  &  16.254  &  16.126  &  15.990  &  15.846  &  15.694   \\
 7.2  &  16.952  &  16.954  &  16.942  &  16.901  &  16.840  &  16.740  &  16.625  &  16.501  &  16.368  &  16.227  &  16.077   \\
\hline
\end{tabular}
} 
\caption {$\log\sigma_0$ for density-dependent composition at $B_{12}=100$} \label{tab:sigma0_b13_comp}
{\tiny
\begin{tabular}{ cccccccccccc }
\hline
  $\log\rho$  &  $\log T$=-1.0  &  $\log T$=-0.8  &  $\log T$=-0.6  &  $\log T$=-0.4  &  $\log T$=-0.2  &  $\log T$=0.0  &  $\log T$=0.2  &  $\log T$=0.4  &  $\log T$=0.6  &  $\log T$=0.8  &  $\log T$=1.0  \\
  \hline
6.0  &  13.171  &  13.068  &  12.948  &  12.801  &  12.654  &  12.512  &  12.370  &  12.226  &  12.078  &  11.925  &  11.765   \\
 6.2  &  13.475  &  13.398  &  13.297  &  13.163  &  13.024  &  12.886  &  12.748  &  12.607  &  12.461  &  12.309  &  12.151   \\
 6.4  &  13.771  &  13.718  &  13.641  &  13.521  &  13.392  &  13.259  &  13.124  &  12.986  &  12.842  &  12.692  &  12.536   \\
 6.6  &  14.061  &  14.031  &  13.961  &  13.875  &  13.757  &  13.630  &  13.500  &  13.364  &  13.223  &  13.075  &  12.920   \\
 6.8  &  14.350  &  14.336  &  14.290  &  14.224  &  14.118  &  13.999  &  13.874  &  13.742  &  13.603  &  13.457  &  13.304   \\
 7.0  &  14.657  &  14.652  &  14.625  &  14.563  &  14.486  &  14.374  &  14.254  &  14.126  &  13.990  &  13.846  &  13.694   \\
 7.2  &  14.952  &  14.954  &  14.942  &  14.901  &  14.840  &  14.740  &  14.625  &  14.501  &  14.368  &  14.227  &  14.077   \\
\hline
\end{tabular}
} 
\end{center}
\end{table}
\begin{table}
\begin{center}
\caption {$\log\sigma_1$ for density-dependent composition at $B_{12}=1$} \label{tab:sigma0_b12_comp}
{\tiny
\begin{tabular}{ cccccccccccc }
\hline
  $\log\rho$  &  $\log T$=-1.0  &  $\log T$=-0.8  &  $\log T$=-0.6  &  $\log T$=-0.4  &  $\log T$=-0.2  &  $\log T$=0.0  &  $\log T$=0.2  &  $\log T$=0.4  &  $\log T$=0.6  &  $\log T$=0.8  &  $\log T$=1.0  \\
  \hline
6.0  &  18.600  &  18.601  &  18.601  &  18.601  &  18.597  &  18.596  &  18.597  &  18.597  &  18.597  &  18.597  &  18.597   \\
 6.2  &  18.800  &  18.800  &  18.801  &  18.801  &  18.798  &  18.796  &  18.797  &  18.797  &  18.797  &  18.797  &  18.797   \\
 6.4  &  18.999  &  19.000  &  19.000  &  19.000  &  18.999  &  18.996  &  18.996  &  18.997  &  18.997  &  18.997  &  18.997   \\
 6.6  &  19.199  &  19.199  &  19.199  &  19.200  &  19.200  &  19.196  &  19.196  &  19.197  &  19.197  &  19.197  &  19.197   \\
 6.8  &  19.398  &  19.398  &  19.398  &  19.399  &  19.399  &  19.396  &  19.396  &  19.396  &  19.397  &  19.397  &  19.397   \\
 7.0  &  19.583  &  19.583  &  19.583  &  19.585  &  19.586  &  19.582  &  19.583  &  19.584  &  19.584  &  19.584  &  19.585   \\
 7.2  &  19.780  &  19.780  &  19.780  &  19.781  &  19.783  &  19.783  &  19.782  &  19.783  &  19.784  &  19.784  &  19.784   \\
\hline
\end{tabular}
} 
\caption {$\log\sigma_1$ for density-dependent composition at $B_{12}=10$} \label{tab:sigma0_b13_comp}
{\tiny
\begin{tabular}{ cccccccccccc }
\hline
  $\log\rho$  &  $\log T$=-1.0  &  $\log T$=-0.8  &  $\log T$=-0.6  &  $\log T$=-0.4  &  $\log T$=-0.2  &  $\log T$=0.0  &  $\log T$=0.2  &  $\log T$=0.4  &  $\log T$=0.6  &  $\log T$=0.8  &  $\log T$=1.0  \\
  \hline
6.0  &  17.601  &  17.601  &  17.601  &  17.601  &  17.597  &  17.596  &  17.597  &  17.597  &  17.597  &  17.597  &  17.597   \\
 6.2  &  17.801  &  17.801  &  17.801  &  17.801  &  17.799  &  17.796  &  17.797  &  17.797  &  17.797  &  17.797  &  17.797   \\
 6.4  &  18.001  &  18.001  &  18.001  &  18.001  &  18.000  &  17.996  &  17.997  &  17.997  &  17.997  &  17.997  &  17.997   \\
 6.6  &  18.201  &  18.201  &  18.201  &  18.201  &  18.201  &  18.196  &  18.197  &  18.197  &  18.197  &  18.197  &  18.197   \\
 6.8  &  18.401  &  18.401  &  18.401  &  18.401  &  18.401  &  18.396  &  18.397  &  18.397  &  18.397  &  18.397  &  18.397   \\
 7.0  &  18.589  &  18.589  &  18.589  &  18.589  &  18.589  &  18.584  &  18.585  &  18.585  &  18.585  &  18.585  &  18.585   \\
 7.2  &  18.789  &  18.789  &  18.789  &  18.789  &  18.789  &  18.787  &  18.785  &  18.785  &  18.785  &  18.785  &  18.785   \\
\hline
\end{tabular}
} 
\end{center}
\end{table}
\begin{table}
\begin{center}
\caption {$\log\sigma_1$ for density-dependent composition at $B_{12}=100$} \label{tab:sigma0_b14_comp}
{\tiny
\begin{tabular}{ cccccccccccc }
\hline
  $\log\rho$  &  $\log T$=-1.0  &  $\log T$=-0.8  &  $\log T$=-0.6  &  $\log T$=-0.4  &  $\log T$=-0.2  &  $\log T$=0.0  &  $\log T$=0.2  &  $\log T$=0.4  &  $\log T$=0.6  &  $\log T$=0.8  &  $\log T$=1.0  \\
  \hline
6.0  &  16.601  &  16.601  &  16.601  &  16.601  &  16.597  &  16.596  &  16.597  &  16.597  &  16.597  &  16.597  &  16.597   \\
 6.2  &  16.801  &  16.801  &  16.801  &  16.801  &  16.799  &  16.796  &  16.797  &  16.797  &  16.797  &  16.797  &  16.797   \\
 6.4  &  17.001  &  17.001  &  17.001  &  17.001  &  17.000  &  16.996  &  16.997  &  16.997  &  16.997  &  16.997  &  16.997   \\
 6.6  &  17.201  &  17.201  &  17.201  &  17.201  &  17.201  &  17.196  &  17.197  &  17.197  &  17.197  &  17.197  &  17.197   \\
 6.8  &  17.401  &  17.401  &  17.401  &  17.401  &  17.401  &  17.396  &  17.397  &  17.397  &  17.397  &  17.397  &  17.397   \\
 7.0  &  17.589  &  17.589  &  17.589  &  17.589  &  17.589  &  17.584  &  17.585  &  17.585  &  17.585  &  17.585  &  17.585   \\
 7.2  &  17.789  &  17.789  &  17.789  &  17.789  &  17.789  &  17.787  &  17.785  &  17.785  &  17.785  &  17.785  &  17.785   \\
\hline
\end{tabular}
} 
\end{center}
\end{table}
\end{widetext}

\end{document}